\pdfoutput=1

\documentclass[11pt,twoside,a4paper,cmspaper,final,collab]{cms-tdr}
\def\svnVersion{c829066}\def\svnDate{2020/06/26}\def\cmsCernNoTag{CERN-EP-2020-012}\def\cmsCernDate{\today}\def\cmsMessage{'Published in the Journal of High Energy Physics as \href{http://dx.doi.org/10.1007/JHEP08(2020)051}{\doi{10.1007/JHEP08(2020)051}.}'}

\begin{document}\cmsNoteHeader{TOP-19-004}

\hyphenation{had-ron-i-za-tion}
\hyphenation{cal-or-i-me-ter}
\hyphenation{de-vices}

\renewcommand{\cmsPubBlock}{\begin{tabular}[t]{@{}r@{}l} &CMS \cmsSTYLE\ \cmsNUMBER\\ &ATLAS TOPQ-2018-002\\ \end{tabular}}
\renewcommand{\cmslogo}{\includegraphics[height=2cm]{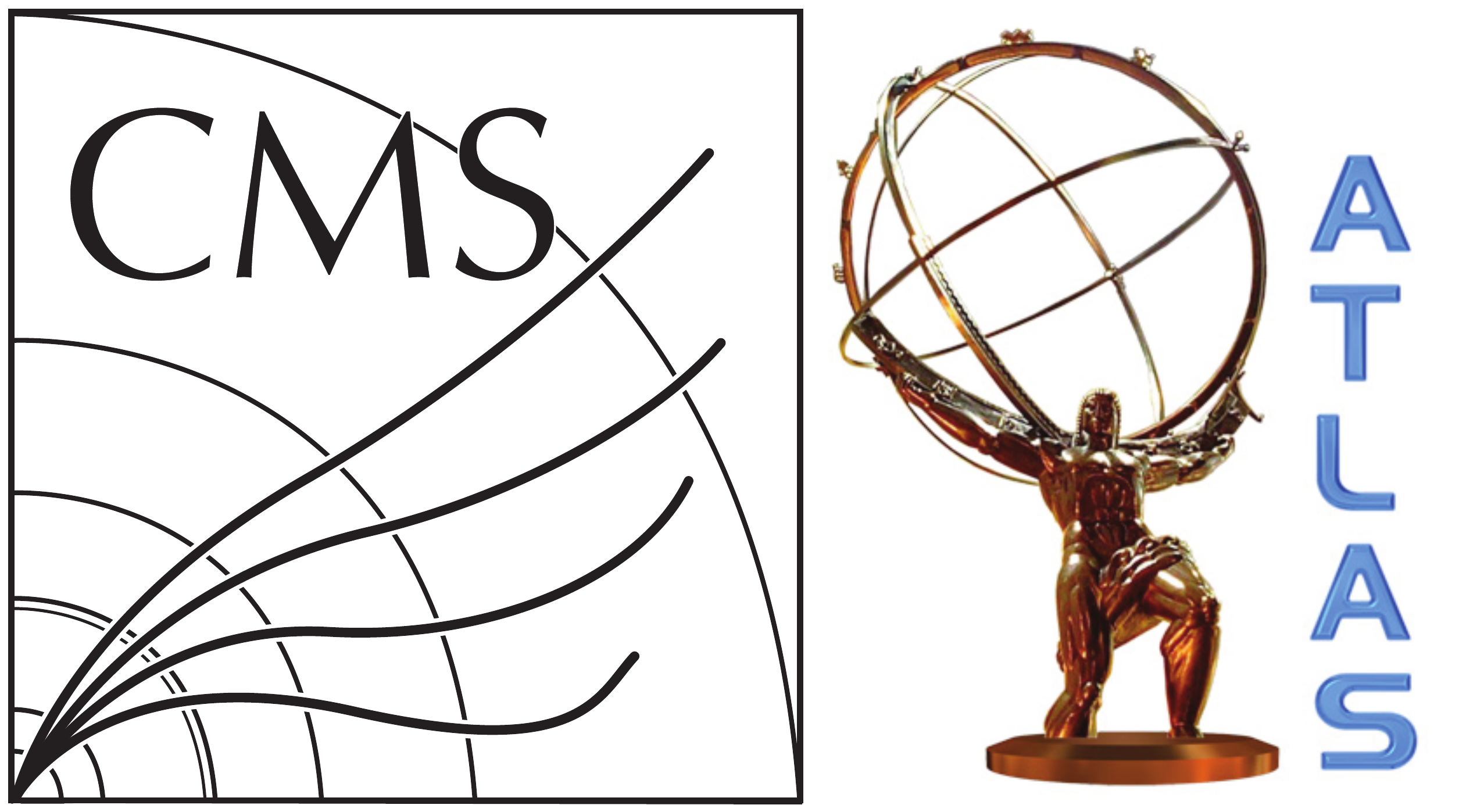}}
\renewcommand{\cmsCollabName}{The CMS and  ATLAS Collaborations}
\renewcommand{\cmsCernNoTag}{CERN-EP-2020-012}
\renewcommand{\cmsTag}{\begin{tabular}[t]{@{}r@{}l} &CMS  \cmsNUMBER\\ &ATLAS-TOPQ-2018-002\\ \end{tabular}}
\renewcommand{\appMsg}{See Appendices~\ref{app:collab} and \ref{app:ATLAScollab} for the lists of collaboration members}
\renewcommand{\cmsCopyright}{\copyright\,\the\year\ CERN for the benefit of the CMS and ATLAS Collaborations.} 
  
\providecommand{\cmsTable}[1]{\resizebox{\textwidth}{!}{#1}}
\newlength\cmsTabSkip\setlength{\cmsTabSkip}{1ex}
\renewcommand{\NA}{\ensuremath{\textit{n.a.}}}
\newcommand{\rhoZL}[1]{\ensuremath{\rho_{#1}(\FZ,\FL)}}
\newcommand{\rhoZZ}[1]{\ensuremath{\rho_{#1}(\FZ,\FZ)}}
\newcommand{\rhoLL}[1]{\ensuremath{\rho_{#1}(\FL,\FL)}}
\newcommand{\rhoIJ}[1]{\ensuremath{\rho_{#1}(\Fi,\Fj)}}
\newcommand{\Fi}{\ensuremath{F_{\text{i}}}}
\newcommand{\Fj}{\ensuremath{F_{\text{j}}}}
\newcommand{\Fjc}{\ensuremath{F_{\text{j}}^{\text{CMS}}}}
\newcommand{\FZ}{\ensuremath{F_{\text{0}}}}
\newcommand{\FL}{\ensuremath{F_{\text{L}}}}
\newcommand{\FLc}{\ensuremath{F_{\text{L}}^{\text{CMS}}}}
\newcommand{\FR}{\ensuremath{F_{\text{R}}}}
\newcommand{\costs}{\ensuremath{\cos \theta^{*}}}
\newcommand{\wbos} {\PW boson\xspace}
\newcommand{\wboss} {\PW bosons\xspace}
\newcommand{\wtb}{\ensuremath{\PQt\PW\PQb}\xspace}
\newcommand{\tq}{top quark\xspace}

\newcommand{\combFZ}{0.693}
\newcommand{\combFZStat}{0.009}
\newcommand{\combFZSyst}{0.011}

\newcommand{\combFL}{0.315}
\newcommand{\combFLStat}{0.006}
\newcommand{\combFLSyst}{0.009}

\newcommand{\combFR}{$-0.008$}
\newcommand{\combFRstat}{0.005}
\newcommand{\combFRsys}{0.006}

\newcommand{\combFRLim}{0.007}

\newcommand{\combChi}{4.3}
\newcommand{\combProb}{64\%}
\newcommand{\combRho}{$-0.85$}

\cmsNoteHeader{TOP-19-004} 

\title{Combination of the \texorpdfstring{\wbos}{W boson} polarization measurements in top quark decays using ATLAS and CMS data at \texorpdfstring{$\sqrt{s} = 8\TeV$}{sqrt(s) = 8 TeV}}

\date{\today}

\abstract{
The combination of measurements of the \PW boson polarization in top quark decays performed by the ATLAS and CMS Collaborations is presented. The measurements are based on proton-proton collision data produced at the LHC at a centre-of-mass energy of 8\TeV, and corresponding to an integrated luminosity of about 20\fbinv for each experiment. The measurements used events containing one lepton and having different jet multiplicities in the final state. The results are quoted as fractions of \PW bosons with longitudinal (\FZ), left-handed (\FL), or right-handed (\FR) polarizations. The resulting combined measurements of the polarization fractions are $\FZ = 0.693 \pm 0.014$ and $\FL = 0.315 \pm 0.011$. The fraction \FR\ is calculated from the unitarity constraint to be $\FR = -0.008 \pm 0.007$.
These results are in agreement with the standard model predictions at next-to-next-to-leading order in perturbative quantum chromodynamics and represent an improvement in precision of 25 (29)\% for \FZ\ (\FL) with respect to the most precise single measurement.
A limit on anomalous right-handed vector ($V_{\text{R}}$), and left- and right-handed tensor ($g_{\text{L}},g_{\text{R}}$) $\PQt\PW\PQb$ couplings is set while fixing all others to their standard model values. The allowed regions are $[-0.11, 0.16]$ for $V_{\text{R}}$, $[-0.08, 0.05]$ for $g_{\text{L}}$, and $[-0.04, 0.02]$ for $g_{\text{R}}$, at 95\% confidence level. Limits on the corresponding Wilson coefficients are also derived.
}

\hypersetup{
pdfauthor={CMS Collaboration},
pdftitle={Combination of the W boson polarization measurements in top quark decays using ATLAS and CMS data at sqrt{s} = 8 TeV},
pdfsubject={CMS,ATLAS,LHCTopWG},
pdfkeywords={CMS, ATLAS, top properties, helicity fractions, LHC combination}}
\maketitle

\section{Introduction}

The large number of top quarks produced at the CERN LHC provides an excellent laboratory for the study of their production and decay properties. Precise predictions of some of these properties are available in the standard model (SM) of particle physics, and are tested through detailed comparisons to data. Potential deviations between data and predictions could reveal important information on the existence of new physics beyond the SM. The properties of the \tq decay vertex \wtb are governed by the structure of the weak interaction. In the SM, this interaction has a $V-A$ structure, where $V$ and $A$ refer to the vector and axial-vector components of the weak current. This structure, along with the masses of the particles involved, determines the fractions of \wboss\ with longitudinal (\FZ), left-handed (\FL), and right-handed (\FR) polarizations, referred to as \emph{polarization fractions}. Theoretical calculations at next-to-next-to-leading order (NNLO) in perturbative quantum chromodynamics (QCD) predict the fractions to be \mbox{$\FZ=0.687 \pm 0.005$}, \mbox{$\FL=0.311 \pm 0.005$}, and \mbox{$\FR=0.0017 \pm 0.0001$}~\cite{Czarnecki:2010gb}, assuming a \tq mass of $172.8 \pm 1.3\GeV$. Thus, the SM predictions can be tested in high-precision measurements of the polarization fractions, and potential new physics processes that modify the structure of the \wtb vertex can be probed.

Experimentally, polarization fractions can be measured in events containing top quarks, using the kinematic properties of its decay products.

For semileptonically decaying top quarks, \ie~$\PQt\to \PW (\to \ell \Pgn)\PQb$ (with lepton $\ell$ = electron, muon, or $\Pgt$), the polarization angle $\theta^{*}$ is defined as the angle between the direction of the charged lepton and the reversed direction of the \PQb quark, both in the rest frame of the \wbos. The distribution of the variable \costs\ is particularly sensitive to the polarization fractions. The differential decay rate is given by
\begin{linenomath}
\begin{equation}
\frac{1}{\Gamma} \frac{\rd \Gamma}{\rd \costs} =    \frac{3}{4} \left( 1-\cos^{2} \theta^{*} \right) \, \FZ + \frac{3}{8} \left(1-\costs \right)^2 \, \FL + \frac{3}{8} \left(1 + \costs\right)^{2} \, \FR.
\label{eq:costh}
\end{equation}
\end{linenomath}
In a similar way, $\theta^{*}$ can be defined for the hadronically decaying top quarks, \ie~$\PQt\to \PW (\to \PQq'\PAQq)\PQb$, by replacing the charged lepton with the down-type quark ($\PQq'$). In the measurements used in this paper, only angles from top quarks decaying semileptonically to electrons or muons are considered. Imposing a unitarity constraint between the three polarization fractions, $\FZ+\FL+\FR=1$, results in two independent observables.

The \wbos polarization fractions have been measured in proton-antiproton collisions by the CDF and D0~experiments~\cite{TevatronCombin} at a centre-of-mass energy of 1.96\TeV with experimental uncertainties of 10--15\% in \FZ~and \FL. The ATLAS and CMS Collaborations have performed measurements at the LHC in proton-proton ($\Pp\Pp$) collisions at $\sqrt{s}=7$~\cite{TOPQ-2011-10,CMS-TOP-11-020} and 8~\cite{TOPQ-2016-02, CMS-TOP-13-008, CMS-TOP-12-020}\TeV, reaching a precision in \FZ~and \FL~of 3--5\%.
All measurements are in agreement with the SM NNLO predictions within their experimental uncertainties. However, these experimental uncertainties are larger than those of the current theoretical predictions, which are less than 2\%. Improving the experimental precision motivates the combination of the ATLAS and CMS measurements: combining measurements based on independent data sets reduces the statistical uncertainty, while the overall uncertainty can be further decreased by exploiting the differences in experimental systematic effects stemming from the use of the two detectors and different analysis methods.

This paper describes the combination of the \wbos polarization fractions measured by the ATLAS and CMS Collaborations based on data collected at $\sqrt{s}=8\TeV$, in final states enhanced in \tq pair (\ttbar)~\cite{TOPQ-2016-02, CMS-TOP-13-008} and single top quark~\cite{CMS-TOP-12-020} production processes. The paper is structured as follows: the measurements included in the combination are briefly described in Section~\ref{sec:measurements}. Section~\ref{sec:systematics} lists the sources of systematic uncertainty considered in the input measurements. The correlations between the measured values included in this combination are categorized in Section~\ref{sec:sec4}, and presented for each source of systematic uncertainty. In Section~\ref{sec:results}, the results of the combination and their interpretation in terms of new physics using the effective field theory approach are described. A summary and conclusions are presented in Section~\ref{sec:conclusion}.

\section{The ATLAS and CMS measurements}
\label{sec:measurements}

Three measurements of the \wbos polarization in the \tq decay from \tq pair production events in the $\ell+$jets channel and one from events with a single \tq signature are the four input measurements in this combination. The measurements based on \ttbar production events were performed by the ATLAS~\cite{TOPQ-2016-02} and CMS~\cite{CMS-TOP-13-008} experiments, where the latter was separated in electron and muon channels. The measurement from events with a single \tq signature was performed by the CMS~\cite{CMS-TOP-12-020} experiment.

The measurements were based on $\Pp\Pp$ collision data at $\sqrt{s}= 8$\TeV, corresponding to integrated luminosities of 20.2 and 19.7\fbinv for the ATLAS and CMS experiments, respectively. The 7\TeV measurements~\cite{TOPQ-2011-10,CMS-TOP-11-020} are not included in this combination since they are based on smaller data sets, and, having relatively large systematic uncertainties, their contribution to the combination is expected to be  marginal. All measurements were based on fits where the polarization fractions were adjusted to describe the observed \costs\ distributions of the semileptonically decaying top quark, taking into account the SM predictions for the backgrounds. These measurements are summarized in the rest of the section. Detailed descriptions of the ATLAS and CMS detectors can be found elsewhere~\cite{atlasDet,cmsDet}.

\subsection{The ATLAS measurement}
\label{sec:altasmeas}

The contributing input from the ATLAS experiment to this combination is described in Ref.~\cite{TOPQ-2016-02} and denoted ``ATLAS'' in the following.
In this measurement, the event selection was defined to efficiently select events from \tq pair decays in the $\ell+$jets channel, \ie exactly one reconstructed electron or muon and at least four jets, of which at least two were tagged as $\PQb$ jets, and  minimizing background contributions, \eg from $\PW$/$\PZ$+jets and multijet productions. The latter corresponds to events including jets misidentified as leptons, or non-prompt leptons from hadron decay passing the $\ell$+jets selection.
 The \ttbar system was fully reconstructed via a kinematic likelihood fit technique~\cite{Erdmann:2013rxa}, which maps the four decay quarks  (two $\PQb$ quarks and two light quarks from the \wbos decay) to four reconstructed jets, utilising Breit--Wigner distributions for the \wbos and top quark masses, as well as transfer functions to map the reconstructed jet and lepton energies to the parton or true lepton level, respectively.

The \wbos polarization was measured in the single-lepton channels from \ttbar events using a template fit method. Dedicated \ttbar templates of the \costs\ distribution for each polarization configuration were produced by reweighting the simulated SM \ttbar events. Additional templates for background processes were also produced.

The templates were fit to the \costs\ distribution in data using different templates for the electron and muon channels, via a binned likelihood fit as:
\begin{linenomath}
\begin{equation}
{\mathcal{L}}=\prod\limits_{k=1}^{n_{\text{bins}}} \frac{N_\text{exp}(k)^{~N_\text{data}(k)}}{\Bigl[ N_\text{data}(k)\Bigr]!}~\exp{[-N_\text{exp}(k)]}
\prod\limits_{j=1}^{n_{\text{bkg}}}\frac{1}{\sqrt{2\pi}\sigma_{\text{bkg}, j}}\exp\left({\frac{-(N_{\text{bkg}, j}-\hat N_{\text{bkg}, j})^2}{2\sigma ^{2}_{\text{bkg}, j}}}\right),
\label{eq:LHFit}
\end{equation}
\end{linenomath}
where $N_{\text{data}}(k)$ and $N_{\text{exp}}(k)$ represented the number of observed and the total number of expected events (sum of signal and background events) in each bin $k$ of the \costs\ distribution, respectively. The number of events for each background source $j$ is represented by $N_{\text{bkg}, j}$. The expected number of events for each background source $j$, $\hat N_{\text{bkg}, j}$, and the uncertainties in the normalization of the background events, $\sigma_{\text{bkg}, j}$, were used to constrain the fit. Therefore, the uncertainties in the polarization fractions obtained from the fit included both the statistical and systematic uncertainties in the background normalizations. The final result was obtained by a simultaneous fit of the electron and muon channel templates to the data. A common parameter was used to scale each of the backgrounds in the electron and muon channel in a fully correlated manner, except in the case of the nonprompt-lepton background for which two separate, uncorrelated, parameters were used. The contribution from $\PW$+jets events was split into different quark flavour samples and scaled by the calibration factors derived from sidebands in data. These procedures were found to cover the corresponding shape uncertainties in the nonprompt-lepton and $\PW$+jets contributions. The uncertainty in the shape of the contributions from single top quark and diboson events was found to be negligible.

\subsection{The CMS measurements}
\label{sec::cms}
Three CMS measurements contribute to this combination.
The results presented in Ref.~\cite{CMS-TOP-13-008} used similar final states to those in ATLAS: one lepton and four or more jets, of which at least two were tagged as $\PQb$ jets. The \ttbar system was fully reconstructed using a constrained kinematic fit. The unmeasured longitudinal momentum of the neutrino was inferred by the kinematic constraints.

The measurement was performed by maximizing the binned Poisson likelihood function,
\begin{linenomath}
 \begin{equation}
\mathcal{L} = \prod\limits_{k=1}^{n_{\text{bins}}} \frac{N_\text{exp}(k)^{\displaystyle~N_\text{data}(k)}}{\Bigl[ N_\text{data}(k)\Bigr]!}~\exp{[-N_\text{exp}(k)]},
\label{eq:lagrang}
 \end{equation}
\end{linenomath}
\noindent
where ${N_\text{data}}(k)$ is the number of observed events in each bin $k$ of the reconstructed \costs~distribution,
and ${ N_\text{exp}}(k)$ is the number of expected events from Monte Carlo (MC) simulation for a given polarization configuration $\vec{F} \equiv  (\FZ ,\FL, \FR)$, including signal and background events.
During each step of the maximization, ${ N_\text{exp}}(k)$ was modified for different values of the polarization fractions $\vec{F}$ using a reweighting procedure based on Eq.~(\ref{eq:costh}). Weights are applied to the events at the  generated level, so that the $\cos\theta^*$ distribution generated according to Eq.~(\ref{eq:costh}) corresponds to alternative values of  $\vec{F}$. Backgrounds that did not involve a top quark did not change ${N_\text{exp}}(k)$ for different values of $\vec{F}$. The ATLAS and CMS measurements considered the variations on ${N_\text{exp}}(k)$ coming from all top quark events passing the selection, either $\ell$+jets or non-$\ell$+jets, including $\Pgt$+jets and dilepton \ttbar processes. In addition, the CMS analyses took into account the variations arising from single top quark processes, which were treated as a background in the ATLAS measurement. The normalization of the \ttbar process was left free in the fit.

In order to allow a more detailed account of the correlations with the other measurements, the two lepton channels, $\Pe$+jets and $\Pgm$+jets, enter the combination as two separate measurements, referred to as ``CMS ($\Pe$+jets)'' and ``CMS ($\Pgm$+jets)'' throughout this paper, respectively. In the ATLAS measurement, the fractions were obtained simultaneously using the events from the two channels, therefore this separation is not available.

The third CMS input~\cite{CMS-TOP-12-020} included in the combination used a final state targeting $t$-channel single top quark topologies instead of \ttbar events. The event selection required exactly one electron or muon, and exactly two jets, one of which was tagged as a $\PQb$ jet. This selection is orthogonal to that of  the CMS ($\Pe$+jets) and CMS ($\Pgm$+jets) analyses, making the three of them statistically independent. Nevertheless, while the expected amount of selected $t$-channel single top quark events corresponded to only about 13\% of the sample, the expected contribution from the \ttbar process amounted to about 35\%, and needed to be taken into account as part of the signal. The largest background came from the $\PW$+jets process. This contribution was fully estimated from data, and corresponded to about 36\% of the selected sample. Other processes, such as multijet and $\PZ$+jets production, accounted for the remaining 16\% of the sample.

The fitting procedure applied in Ref.~\cite{CMS-TOP-13-008} was slightly modified for the single top quark topology measurement. In this case, because of the different background composition with respect to the \ttbar analysis, the normalizations of the single top quark and \ttbar processes were fixed according to their predicted cross section values. On the other hand, the normalization of the $\PW$+jets sample was left free in the fit  to be adjusted simultaneously with the \FZ\ and \FL\ fractions, and treated independently in the $\Pe$+jets and $\Pgm$+jets channels.
Moreover, the fractions were extracted by maximizing a combined likelihood function, constructed from the two likelihood functions of the electron and muon channels, taking into account
the correlations between them. Therefore, although based on two single-lepton channels, this measurement contributes to the combination as one single input, denoted as ``CMS (single top)'' in the following.

\subsection{The \texorpdfstring{$\PW$ boson polarization values from the input measurements}{}}
\label{sec:mod_measurements}

The polarization fractions from the input measurements before applying the modifications concerning the combination (as discussed in Section~\ref{sec:systematics}), and their uncertainties are summarized in Table~\ref{tab:measurements}. The first quoted uncertainty in the ATLAS measurement includes  the statistical uncertainties and uncertainties in the background determination, and the second uncertainty refers to the remaining systematic uncertainty. For CMS measurements, the first uncertainty is statistical, while the second is the total systematic uncertainty, including that on background determination. 

In order to harmonize the treatment of the systematic uncertainties evaluation across the input measurements, some of them are modified before performing the combination process. The following modifications are applied (as detailed in Section~\ref{sec:systematics}):
\begin{itemize}
    \item The uncertainty values in the ATLAS measurement are symmetrized.

    \item The \ttbar modelling uncertainties in the CMS ($\Pe$+jets) and CMS ($\Pgm$+jets) measurements are recalculated without the contributions from the limited number of events in the samples used to estimate them.

    \item The uncertainty due to the top quark mass used in the ATLAS measurement is increased from a variation of $\pm 0.7\GeV$ to $\pm 1.0\GeV$.
\end{itemize}

\renewcommand{\arraystretch}{1.2}
\begin{table}[ht!]
\centering
\topcaption{Summary of the published ATLAS and CMS measurements for 8\TeV data. The first quoted uncertainty in the ATLAS measurement includes statistical uncertainties and uncertainties in the background determination, and the second uncertainty refers to the remaining systematic contribution. For CMS measurements, the first uncertainty is statistical while the second is the total systematic uncertainty, including that on background determination.
\label{tab:measurements}}

\begin{tabular}{lccc}
\hline
Measurement &  \FZ & \FL & \FR\\
\hline
ATLAS ($\ell$+jets)  & $0.709 \pm 0.012 \pm 0.015$ & $ 0.299\pm 0.008 \pm 0.013$ & $ -0.008 \pm 0.006 \pm 0.012$  \\
CMS ($\Pe$+jets) & $0.705 \pm 0.013 \pm 0.037$ & $0.304 \pm     0.009 \pm 0.020$ & $ -0.009 \pm 0.005 \pm 0.021$  \\
CMS ($\Pgm$+jets)   & $0.685 \pm 0.013 \pm 0.024$ & $0.328 \pm 0.009 \pm 0.014$ & $-0.013 \pm 0.005 \pm 0.017$ \\
CMS (single top) & $0.720 \pm 0.039 \pm 0.037$ & $0.298 \pm 0.028 \pm 0.032$ & $-0.018 \pm 0.019 \pm 0.011$ \\
\hline
\end{tabular}
\end{table}

\section{Sources of systematic uncertainty}
\label{sec:systematics}
The effects of various systematic uncertainties on the input results were studied individually for each measurement. In the ATLAS measurement, the impact of systematic uncertainties was evaluated with alternative pseudo-data distributions built from the altered signal and background contributions. The alternative pseudo-data distributions were produced by varying each source of systematic uncertainty by one standard deviation ($\pm 1\sigma$). The CMS measurements also used pseudo-data to estimate the uncertainties due to parton distribution functions (PDFs), size of the simulated samples, and single top quark analysis specific uncertainties. The other uncertainties were estimated by replacing the nominal sample with alternative samples containing simulated events modified according to each of the systematic variations, and repeating the fit.

As the algorithm used to perform the combination accepts only symmetric uncertainties (more details in Section~\ref{sec:results}), the uncertainties in the ATLAS measurement are symmetrized by assigning the average uncertainty value between the up and down variations in each uncertainty source. A test is performed by replacing the average uncertainty value with the largest shift among the up and down variations. No variation in the combination results is observed, \ie the central values of the polarization fractions, combination uncertainty, and total correlation remain unchanged. In addition, common uncertainty categories are established by merging and regrouping various uncertainties in each individual input measurement.

In the following, the categorization of the systematic uncertainties considered for the combination is presented. The categories, assumed to be independent from each other, comprise sources of uncertainties that have similar origins, easing the treatment of correlations discussed in Section~\ref{sec:sec4}.

\subsection{Limited size of the data and simulated samples, backgrounds, and integrated luminosity}

\noindent {\textit{Statistical uncertainty, background determination, and integrated luminosity (stat+bkg):}}
The uncertainties in the ATLAS measurement from the fit included both the statistical uncertainty in the data and the systematic uncertainty in the background normalizations via priors for the background yields. The shape of the multijet processes was determined from data, while for the other background events it was fully determined from simulation. The impact of the 1.9\% integrated luminosity uncertainty~\cite{DAPR-2013-01} was found to be negligible because of the background normalization treatment in the fit.

In the CMS measurements, the uncertainties in the expected backgrounds included shape and normalization effects, and were estimated by varying them separately within their uncertainties and repeating the measurement. The multijet background in all CMS measurements as well as the normalization of the $\PW$+jets contribution in the CMS (single top) case were derived exclusively from data.
All other background processes, as well as \ttbar, and single top quark processes in the CMS (single top) measurement were estimated using simulation, normalized to the  integrated luminosity of the data samples. These were affected by the uncertainties in their predicted cross sections, and the integrated luminosity determination. The CMS integrated luminosity uncertainty of  2.6\%~\cite{CMS-PAS-LUM-13-001} had a sizeable effect only on the CMS (single top) measurement. 

\noindent {\textit{Size of simulated samples:}}
This category accounts for the limited number of simulated events for the nominal samples in all input measurements. Both ATLAS and CMS evaluated this uncertainty by performing pseudo-experiments. In the CMS ($\Pe$+jets) and CMS ($\Pgm$+jets) measurements, the limited number of simulated events was also considered for the \ttbar samples used for the estimation of the modelling uncertainties. In order to perform a consistent combination, the \ttbar modelling uncertainties in the CMS ($\Pe$+jets) and CMS ($\Pgm$+jets) measurements are recalculated without the contributions from the limited number of events in the samples used to estimate them. The impact of this modification on the relative uncertainty in the measurements is found to be in the order of $O(10^{-4})$.

\subsection{Detector modelling}
\label{sec:detmod}

\noindent {\textit{Jets:}} In all input measurements in this combination, the same jet clustering algorithm, the anti-\kt algorithm~\cite{Cacciari:2008gp,Cacciari:2011ma}, was used, with the radius parameter R of 0.4 and 0.5 for the ATLAS and CMS experiments, respectively. However, in the ATLAS measurement the jets were built from energy deposits in the calorimeter~\cite{PERF-2014-07}, while in the CMS analyses they were reconstructed from particle-flow~\cite{CMS-PRF-14-001} objects. Thus, the two experiments used different calibration procedures and uncertainties for jets. The following categories comprise various sources of uncertainty related to the reconstruction and energy calibration of jets.

\begin{itemize}

\item \noindent {\textit{ Jet energy scale (JES):}}
The JES uncertainty in the ATLAS and CMS analyses was composed of different uncertainty sources, such as jet flavour dependence, the additional interactions in the same or nearby bunch crossings (pileup), calibrations from $\PZ$+jets or $\gamma$+jets processes, and other components. In general, these components have different level of correlations among the two experiments and have been used to evaluate the total JES correlation (as detailed in Section~\ref{subsec:stability}). The final JES uncertainty used in this combination is quoted in Tables~\ref{tab:uncertainties_ATLAS}--\ref{tab:uncertainties_CMSst} and results from grouping all JES uncertainty components into a single number.

\item \noindent {\textit{ Jet energy resolution (JER):}}
This category includes contributions due to the uncertainties in the modelling of the jet energy resolution. The momenta of the jets in simulation were smeared so that the jet energy resolution in simulation agrees with that in data. Both experiments used a similar method to estimate this uncertainty.

\item \noindent {\textit{Jet vertex fraction (JVF):}}
To suppress jets from pileup, in the ATLAS measurement jets were required to fulfil the JVF criterion. The corresponding uncertainty was evaluated in the measurement by changing the nominal JVF cutoff value and repeating the measurement~\cite{ATLAS-CONF-2013-083}.
In the CMS measurements, pileup events were removed at the event reconstruction level with the particle-flow algorithm. In this case, uncertainties in jet reconstruction due to pileup were covered by the JES and pileup categories, and are not added as a separate source.

\item \noindent {\textit{Jet reconstruction efficiency:}}
A systematic uncertainty was included in the ATLAS measurement to account for the jet reconstruction efficiency mismatch between simulation and data. In the CMS measurements, this uncertainty is included in the JES uncertainty.
\end{itemize}

\noindent {\textit{Lepton efficiency:}}
For all measurements, this category accounted for the uncertainties in the scale factors used to correct the simulated samples so that the efficiencies for lepton selection, reconstruction, and identification observed in data were well reproduced by the simulation. Since the samples were collected using single-lepton triggers, uncertainties in the trigger efficiencies were also included. All corrections were applied as functions of \pt\ and $\eta$ of the leptons. This uncertainty was found to be negligible for the CMS (single top) measurement, compared to other uncertainties.

\noindent {\textit{$\PQb$ tagging:}}
In this category, uncertainties on the scale factors used to correct the simulation for different efficiencies for tagging jets originating from $\PQb$ quarks (tag) or for those originating from $\PQc$ or light partons  wrongly identified as $\PQb$ jets (mistag) were taken into account. This difference was accounted for by assigning scale factors to the jets, dependent on the \pt and $\eta$ as well as on the flavour of the jet.
In the ATLAS measurement, additionally, an uncertainty was assigned to account for the extrapolation of the $\PQb$ tagging efficiency measurement to the high-\pt region.

\noindent {\textit{Pileup:}}
In both the ATLAS and the CMS analyses, pileup effects were taken into account in the simulation of signal and background events. The distribution of pileup was adjusted to reflect the measured instantaneous luminosities per bunch in data. In the CMS measurements, this uncertainty was estimated by varying the $\Pp\Pp$ cross section used to estimate the number of pileup in data within its uncertainty, and recalculating the weights applied to the simulation. In the ATLAS measurement, the uncertainty in the description of extra energy deposited due to pileup interactions was treated as a separate missing transverse momentum (\ptmiss) scale uncertainty. The impact on the measured \wbos polarization fractions from this uncertainty was found to be negligible, and therefore was not considered.

\subsection{Signal modelling}
\label{sec:sigMod}

\noindent {\textit{Top quark mass:}}
In all four analyses, the effect of the uncertainty in the top quark mass was estimated by repeating the measurements using simulated samples with different input top quark masses for the signal process.
In the ATLAS measurement, this effect was evaluated using an uncertainty of $\pm 0.70\GeV$ in the top quark mass as given by the ATLAS measurement~\cite{TOPQ-2016-03}. In the CMS measurements on the other hand, an uncertainty of $\pm 1.0\GeV$ in the top quark mass was assumed. In order to keep consistency across the various input measurements, this effect in the ATLAS measurement is reestimated using the original estimation method described in Ref.~\cite{TOPQ-2016-02}, accounting for a variation of $\pm 1.0\GeV$ in the top quark mass. The impact of this modification in the ATLAS input result is negligible.

\noindent {\textit{Simulation model choice:}}
The impact of using different MC event generators and their interfaced showering and hadronization models was estimated in all input measurements. In the ATLAS measurement, the impact of the choice of different MC event generators was assessed by comparing events produced by \textsc{Powheg-Box}~\cite{Nason:2004rx,Frixione:2007nw,Frixione:2007vw,Alioli:2009je,Alioli:2010xd} and \MCATNLO~\cite{Frixione:2002ik,Frixione:2003ei,Frixione:2005vw}, both interfaced to \textsc{Herwig}~\cite{Corcella:2000bw} for showering and hadronization. 
To evaluate the impact of the different parton shower and hadronization models, the \textsc{Powheg+Herwig} sample was compared to \textsc{Powheg+Pythia}~\cite{Sjostrand:2006za}.
For the CMS ($\Pe$+jets) and CMS ($\Pgm$+jets) measurements, the uncertainties were estimated by replacing the events produced by \MADGRAPH~\cite{Alwall:2014hca} interfaced with \textsc{Pythia} with \MCATNLO interfaced with the \textsc{Herwig} generator and additionally, varying the kinematic scale used to match jets to partons (matching threshold) by twice and half its central value.
In the CMS (single top) measurement, the uncertainty in the choice of different MC generators was estimated as the difference between the \textsc{Powheg+Pythia} and the \COMPHEP~\cite{Comphep2} generators.

\noindent {\textit{Radiation and scales:}}
In all four analyses, this category represents the uncertainty associated with initial- and final-state radiation (ISR/FSR) estimated using simulated samples of \ttbar events where the renormalization and factorization scales ($\mu_{\mathrm{R}}$ and $\mu_{\mathrm{F}}$) were simultaneously set to twice and half the default value in the matrix element (ME) calculations. In the CMS measurements, the $\mu_{\mathrm{R}}$ and $\mu_{\mathrm{F}}$ in the parton shower were also varied simultaneously to those used in the ME calculations.
However, in the ATLAS measurement, a different set of tuned parameters of the \PYTHIA parton shower with a modified strong coupling \alpS was used to account for low and high radiation to match the variation of scales in the ME calculations. The detailed list of modified parameters is given in Ref.~\cite{Skands:2010ak}.
Furthermore, in the ATLAS measurement the value of the damping parameter ($h_{\text{damp}}$) in \textsc{Powheg-Box} was set to twice the top quark mass for the high-radiation sample.
In addition to changing it in the \ttbar background, the CMS (single top) measurement varied the scales used in the single top quark simulated samples. 

\noindent {\textit{Top quark \pt:}}
In previous CMS analyses of \ttbar events, described \eg in Ref.~\cite{CMS-TOP-12-028}, the shape of the \pt spectrum for top quarks was found to be softer than the predictions from \MADGRAPH\ simulation. The effect of this mismodelling on the CMS ($\Pe$+jets) and CMS ($\Pgm$+jets) measurements was estimated by reweighting the simulated \ttbar sample to describe the data. The difference in the polarization fractions with the default sample to the reweighted sample was taken as a systematic uncertainty.
On the other hand, the top quark \pt distribution did not exhibit, within uncertainties, a significant difference with the predictions in the single top quark enriched phase space, therefore no systematic uncertainty was assigned in the CMS (single top) measurement.
In the ATLAS measurement, this mismodelling was checked to be covered by the simulation model choice uncertainties, and therefore no additional uncertainty for the top quark \pt spectrum was considered.

\noindent {\textit{PDF:}}
{\tolerance=800 
The uncertainty due to the choice of PDFs in all input measurements was evaluated by varying the eigenvalues of different PDF sets following the PDF4LHC recommendations~\cite{Alekhin:2011sk, Botje:2011sn}.
In the ATLAS measurement, the differences between three PDF sets: CT10~\cite{Lai:2010vv}, MSTW2008~\cite{Martin:2009iq}, and NNPDF 2.3~\cite{Ball:2012cx} were taken into account.
 Uncertainties related to the choice of PDF set in the CMS ($\Pe$+jets) and CMS ($\Pgm$+jets) measurements were estimated by replacing CTEQ6L1~\cite{cteq} used to generate the nominal samples, with NNPDF 2.1~\cite{Ball:2011mu} and  MSTW2008. A similar procedure was adopted in the CMS (single top) measurement, where the default CTEQ6.6M~\cite{Nadolsky:2008zw} set was replaced with CT10 instead.
\par}

\noindent {\textit{Single top quark analysis method:}}
In addition to the systematic uncertainties considered for the \ttbar measurement, a few specific uncertainties were included for the CMS (single top) measurement.
For the specific case of single top quark processes, unlike for \ttbar production, the polarization fractions can also be altered at the production level. To study this effect, pseudo-data were generated from samples simulated using \COMPHEP and {\textsc{Single Top}}~\cite{Comphep1} event generators with varied values of the couplings $g_{\text{L}},\ V_{\text{R}}$, and $V_{\text{L}}$ (as described in Section \ref{sec:anomcoupl}) both at the single top quark production and decay, and the polarization fractions values were extracted using the analysis fitting framework. The differences between the generated and fitted values were taken as the systematic uncertainty. Secondly, a small difference in the generated \wbos polarization fraction values was observed for the \ttbar events, simulated with \MADGRAPH, and single top quark events, simulated with \POWHEG. This difference of about 0.01 was taken into account as an uncertainty in the measurement.
Finally, the effect of fixing the signal normalization in the fit was considered.
All these uncertainties are merged into a single uncertainty, referred to as {\textit{Single top method}} in Tables~\ref{tab:uncertainties_ATLAS} to~\ref{tab:uncertainties_LHC}.

In all input measurements, the uncertainty in the modelling of colour reconnection was found to be negligible and therefore was not considered.

\section{Correlations and uncertainties in the ATLAS and CMS measurements}
\label{sec:sec4}

\subsection{Correlations}
\label{sec:correlation}

Four pairs of longitudinal and left-handed polarization fractions from four input measurements, as described in Section~\ref{sec:measurements} are used in the combination. The correlations between the input values are defined taking into account the unitarity relation between the polarization fractions in each measurement and the correlations among the measurements. The groups of correlations are listed in Table~\ref{tab:corr_categories} and defined as follows:

\begin{itemize}
\item \textit{Correlations within the same measurement:}\\
Because of the unitarity constraint, and given that $\FR \approx 0$, the observed values of \FZ\ and \FL\ within one single measurement are usually highly anticorrelated.
In the ATLAS measurement, this correlation is estimated for each systematic uncertainty source from its corresponding covariance matrix. For categories with multiple sources of systematic uncertainty, the sum of the individual covariance matrices is used to calculate the correlation.
In the CMS analyses, this group of correlations is estimated from the covariance propagation of the expression $\FR=1-\FZ-\FL$ as
\begin{linenomath}
\begin{equation}
\rho(\FZ,\FL) = \frac{\sigma^{2}(\FR) - \sigma^{2}(\FZ) -
\sigma^{2}(\FL)}{2\sigma(\FZ)\sigma(\FL)},
\label{eqn:rho}
\end{equation}
\end{linenomath}
where $\sigma(\Fi)$ is the uncertainty in the polarization fraction \Fi , which is directly obtained from the individual measurements. This is done for all sources of systematic uncertainty. For systematic uncertainty categories with multiple sources, \eg `stat+bkg' including statistical uncertainty, background determination, and others, $\sigma^2(\Fi)$ is defined as the quadratic sum of the individual uncertainty sources.

This group of correlations is denoted in this document as $\rho_{\text{ATLAS}}$, $\rho_{\text{CMS}}^{\Pe\text{+jets}}$, $\rho_{\text{CMS}}^{\Pgm\text{+jets}}$, and $\rho_{\text{CMS}}^{\text{st}}$ for the  ATLAS, CMS ($\Pe$+jets), CMS ($\Pgm$+jets), and CMS (single top) measurements, respectively.

\item \textit{ Correlations between measurements within the CMS experiment:}\\
{\tolerance 800
For each source of systematic uncertainty, the correlations between the polarization fractions in the CMS ($\Pe$+jets) and CMS ($\Pgm$+jets) measurements are denoted $\rho_{\text{CMS}}^{\Pe,\Pgm\text{+jets}}(\Fi,\Fj)$, where i and j stand for 0 or L. The correlations between  CMS (single top) and CMS ($\Pe$+jets) are assumed to be the same as those between the CMS (single top) and CMS ($\Pgm$+jets) measurements  for each source of the uncertainty, and are denoted generically $\rho_{\text{CMS}}^{\text{st,}\ell \text{+jets}}(\Fi,\Fj)$. The relations $\rho_{\text{CMS}}(\FZ,\FZ)=\rho_{\text{CMS}}(\FL,\FL)=-\rho_{\text{CMS}}(\FZ,\FL)$ are assumed in all CMS measurements. In this hypothesis, the strong anti-correlation observed for \FZ\ and \FL\ within the same measurement (as described above) is assumed to hold also across different measurements.
\par}
The uncertainties associated with the limited size of the data and simulated samples, and background estimation are assumed to be uncorrelated (as also discussed in Sections~\ref{sec:partialcorrel} and~\ref{subsec:stability}).
The lepton efficiency uncertainty is assumed to be uncorrelated between the CMS ($\Pe$+jets) and CMS ($\Pgm$+jets) measurements, and partially correlated with the CMS (single top) measurement.
All other sources of uncertainty are assumed to be fully correlated.

\item \textit{Correlations between the ATLAS and CMS experiments:}\\
For each source of systematic uncertainty, the correlation between the measured polarization fractions $\Fi$ by the ATLAS and CMS experiments, $\rho(\Fi^{\text{ATLAS}},\Fj^{\text{CMS}})$ is presented by \rhoIJ{\text{LHC}}, where $\rhoZZ{\text{LHC}}=\rhoLL{\text{LHC}}=-\rhoZL{\text{LHC}}$ are assumed.

The uncertainties associated with the detector modelling (except for the JES) as well as  the method-specific uncertainty are assumed to be uncorrelated, \ie~$\rhoZZ{\text{LHC}}=0$.

The uncertainty associated with the radiation and scales, and the JES are assumed to be partially correlated with $\rhoZZ{\text{LHC}}$ estimated to be 0.5 and 0.2, respectively (see Sections~\ref{sec:partialcorrel} and \ref{subsec:stability} for details). All other sources of uncertainty are assumed to be fully correlated, \ie~$\rhoZZ{\text{LHC}}=+1$.
\end{itemize}

\subsection{Correlation choices for the partially correlated uncertainties}
\label{sec:partialcorrel}

Although the correlations  between the measurements are well known for most of the systematic uncertainty sources, some of them, in particular those that are partially correlated, are not very accurately determined.
This section describes how these values are estimated for the combination.
Stability tests are performed to verify the
robustness of the combination against these correlation assumptions, as discussed in Section~\ref{subsec:stability}.

In the CMS measurements, the uncertainties in the background determination (shape and normalization), integrated luminosity, and the statistical uncertainty were estimated independently and grouped into a single uncertainty category (stat+bkg) for coherence with the ATLAS treatment. The major components of the stat+bkg category in the CMS ($\Pe$+jets) and CMS ($\Pgm$+jets) measurements are the uncertainty in the determination of the background events from multijet and $\PW$+jets production.
The former is estimated from data, and therefore uncorrelated  between all CMS measurements, while $\PW$+jets production, as well as the other minor backgrounds are estimated from simulation, and therefore at least partially correlated  between the measurements. For the CMS (single top) case, the major  component of this category is the statistical uncertainty, which is uncorrelated  with the other measurements.
The normalization of $\PW$+jets production, a major background in the CMS (single top) analysis, is estimated from data, and therefore it is uncorrelated to the other CMS measurements. On the other hand, the $\PW$+jets production shape, as well as the modelling of other background event sources and signal events, rely on simulation, which may lead to a nonzero $\rho_{\text{CMS}}^{\text{st,} \ell \text{+jets}}(\Fi,\Fi)$ correlation. Neglecting the small correlations that could arise from the $\PW$+jets production shape and the background modelling from simulation, the values $\rho_{\text{CMS}}^{\Pe,\Pgm\text{+jets}}(\Fi,\Fj)=0$ and $\rho_{\text{CMS}}^{\text{st,}\ell\text{+jets}}(\Fi,\Fi)=0$ are assumed for the combination, and the impact of this assumption is studied via the stability tests.

In all ATLAS and CMS measurements, the JES systematic uncertainty is estimated from different components, which are characterized by different levels of correlations among the two experiments. These components are categorized as fully correlated, such as gluon-initiated jet fragmentation; partially correlated, such as modelling uncertainties from \textit{in situ} techniques,  such as \PZ-jet, \PGg-jet, and multijet balance techniques; and uncorrelated, such as statistical and detector-related uncertainties. These correlations have been evaluated and are described in Ref.~\cite{ATL-PHYS-PUB-2015-049}.
In the ATLAS measurement, the contribution from the uncorrelated (partially correlated) components to the total JES uncertainty is found to be about 70 (20)\%, and the total JES uncertainty is dominated by the uncorrelated jet flavour composition component. In the CMS measurements, because JES uncertainties are small, the breakdown into components was not done. Therefore, assuming a similar JES uncertainty composition between the two experiments, the value of $\rho_{\text{LHC}}(\Fi,\Fi)$ is found to be 0.2.

In the ATLAS and CMS analyses, different approaches were used to estimate the radiation and scales uncertainties, as described in Section~\ref{sec:sigMod}.
In the CMS (single top) measurement, this uncertainty is estimated by varying  the scales $\mu_\text{R}$ and $\mu_\text{F}$ for the simulations of both the \ttbar and the single top quark processes. While the \ttbar component, which is dominant, is fully correlated to the analogous uncertainties in the ATLAS, CMS ($\Pe$+jets), and CMS ($\Pgm$+jets) measurements, the smaller component from the single top quark $\mu_\text{R}$ and $\mu_\text{F}$ scales is uncorrelated  with the other measurements. Since the effects being studied are the same, but the methods are different, the values of $\rho_{\text{LHC}}(\Fi,\Fi)$ and $\rho_{\text{CMS}}^{\text{st,}\ell\text{+jets}}(\Fi,\Fi)$ are not well known, and are assumed to be 0.5 and 1.0, respectively.

\renewcommand{\arraystretch}{1.6}
\begin{table}[b!]
\centering
\topcaption{Summary of the correlation categories considered in the combination. The correlations among the \FL\ measurements are not shown for brevity.
\label{tab:corr_categories}}
\cmsTable{
\begin{tabular}{@{}lcrrrr}
\hline
Measurement   &    & ATLAS         &   CMS ($\Pe$+jets)            & CMS ($\Pgm$+jets)             &    CMS (single top) \\
         & Fraction    & \FZ     & \FZ     & \FZ      & \FZ \\
\hline
 ATLAS & \FZ  & $+1$                 & $\rhoZZ{\text{LHC}}$                     & $\rhoZZ{\text{LHC}}$                     & $\rhoZZ{\text{LHC}}$   \\
 CMS ($\Pe$+jets)   & \FZ  & $\rhoZZ{\text{LHC}}$  & +1                                    & $\rho_{\text{CMS}}^{\Pe,\Pgm\text{+jets}}(\FZ,\FZ)$  & $\rho_{\text{CMS}}^{\text{st,}\ell\text{+jets}}(\FZ,\FZ)$\\
 CMS ($\Pgm$+jets)  & \FZ  & $\rhoZZ{\text{LHC}}$  & $\rho_{\text{CMS}}^{\Pe,\Pgm\text{+jets}}(\FZ,\FZ)$    & $+1$                                    & $\rho_{\text{CMS}}^{\text{st,}\ell\text{+jets}}(\FZ,\FZ)$\\
 CMS (single top)      & \FZ  & $\rhoZZ{\text{LHC}}$  & $\rho_{\text{CMS}}^{\text{st,}\ell\text{+jets}}(\FZ,\FZ)$     & $\rho_{\text{CMS}}^{\text{st,}\ell\text{+jets}}(\FZ,\FZ)$      & $+1$   \\ [\cmsTabSkip]

 ATLAS & \FL  & $\rho_{\text{ATLAS}}(\FZ,\FL)$    & $-\rhoZZ{\text{LHC}}$                    & $-\rhoZZ{\text{LHC}}$                     & $-\rhoZZ{\text{LHC}}$ \\
 CMS ($\Pe$+jets)   & \FL  & $-\rhoZZ{\text{LHC}}$             & $\rho_{\text{CMS}}^{\Pe\text{+jets}}(\FZ,\FL)$        & $-\rho_{\text{CMS}}^{\Pe,\Pgm\text{+jets}}(\FZ,\FZ)$    & $-\rho_{\text{CMS}}^{\text{st,}\ell\text{+jets}}(\FZ,\FZ)$ \\
 CMS ($\Pgm$+jets)  & \FL  & $-\rhoZZ{\text{LHC}}$             & $-\rho_{\text{CMS}}^{\Pe,\Pgm\text{+jets}}(\FZ,\FZ)$   & $\rho_{\text{CMS}}^{\Pgm\text{+jets}}(\FZ,\FL)$       & $-\rho_{\text{CMS}}^{\text{st,}\ell\text{+jets}}(\FZ,\FZ)$ \\
 CMS (single top)      & \FL  & $-\rhoZZ{\text{LHC}}$             & $-\rho_{\text{CMS}}^{\text{st,}\ell\text{+jets}}(\FZ,\FZ)$    & $-\rho_{\text{CMS}}^{\text{st,}\ell\text{+jets}}(\FZ,\FZ)$     & $\rho_{\text{CMS}}^{\text{st}}(\FZ,\FL)$         \\
\hline
\end{tabular}
}
\end{table}

\renewcommand{\arraystretch}{1.1}
\begin{table}[ht!]
\centering
\topcaption{Input correlations across different measurements, as explained in Section~\ref{sec:correlation}. The values stand for the correlations $\rho(\Fi,\Fi)$, with $\text{i}$ being either 0 or L. The correlations of the type $\rho(\FZ,\FL)$ are assumed to be $\rho(\FZ,\FL)=-\rho(\FZ,\FZ)=-\rho(\FL,\FL)$. In case an uncertainty is not applicable, the correlation value is set to zero and marked with an asterisk. The correlations marked with a dagger sign are those that are not precisely determined and checks are performed to test the stability of the results against these assumptions.
}
\label{tab:input_correlations}
\begin{tabular}{@{}llll@{}}
\hline
                              & $\rho_{\text{LHC}}(\Fi,\Fi)$ & $\rho_{\text{CMS}}^{\Pe,\Pgm\text{+jets}}(\Fi,\Fi)$ & $\rho_{\text{CMS}}^{\text{st,}\ell\text{+jets}}(\Fi,\Fi)$ \\
Uncertainty Category  &  &  &  \\
\hline
\multicolumn{4}{@{}l}{\textit{Samples size  and background determination}} \\
\, \, \ Stat+bkg          & 0.0 & $0.0^{\dagger}$ & $0.0^{\dagger}$\\
\, \, \ Size of simulated samples   & 0.0 & 0.0 & 0.0\\ [\cmsTabSkip]
\multicolumn{4}{@{}l}{\textit{Detector modelling}} \\
\, \, \ JES                 & $0.2^{\dagger}$ & 1.0 & 1.0\\
\, \, \ JER                 & 0.0 & 1.0 & 1.0\\
\, \, \ JVF                 & $0.0^{\ast}$ & $0.0^{\ast}$ & $0.0^{\ast}$\\
\, \, \ Jet reconstruction efficiency  & $0.0^{\ast}$ & $0.0^{\ast}$ & $0.0^{\ast}$\\
\, \, \ Lepton efficiency   & 0.0 & 0.0 & 0.0\\
\, \, \ $\PQb$ tagging           & 0.0 & 1.0 & 1.0\\
\, \, \ Pileup              & $0.0^{\ast}$ & 1.0 & 1.0\\ [\cmsTabSkip]
\multicolumn{4}{@{}l}{\textit{Signal modelling}} \\
\, \, \ Top quark mass              & 1.0 & 1.0 & 1.0\\
\, \, \ Simulation model choice     & 1.0 & 1.0 & 1.0\\
\, \, \ Radiation and scales        & $0.5^{\dagger}$ & 1.0 & $1.0^{\dagger}$\\
\, \, \ Top quark \pt               & $0.0^{\ast}$ & 1.0 & $0.0^{\ast}$\\
\, \, \ PDF                         & 1.0 & 1.0 & 1.0\\
\, \, \ Single top  method          & $0.0^{\ast}$ & $0.0^{\ast}$ & $0.0^{\ast}$\\
\hline
\end{tabular}
\end{table}

\subsection{Summary of the uncertainties and correlations of the input measurements}
\label{sec:uncertainties}

For each systematic uncertainty category, the correlations between the measured polarization fractions for the input measurements are given in Table~\ref{tab:input_correlations}. A breakdown of the uncertainties in the input measurements of \FZ\ and \FL\, as well as their correlations, are presented in Tables~\ref{tab:uncertainties_ATLAS}--\ref{tab:uncertainties_CMSst}. The uncertainties are grouped according to the categories listed in Section~\ref{sec:systematics}.

\renewcommand{\arraystretch}{1.1}
\begin{table}[ht!]
\centering
\topcaption{Uncertainties in \FZ, \FL\ and their corresponding correlations from the ATLAS measurement. The uncertainty that is not applicable to this measurement, or which is included in other categories, is indicated by ``\NA''. The line ``Systematic uncertainty'' represents the quadratic sum of all the systematic uncertainty sources except for the uncertainty in the background determination, which is included in the ``Stat+bkg'' category. The quoted correlation values are obtained via the procedures described in Section~\ref{sec:correlation}.
\label{tab:uncertainties_ATLAS}
}
\begin{tabular}{@{}lccc@{}}
\hline
& \multicolumn{3}{@{}c@{}}{ATLAS} \\ 
                      & \FZ & \FL & \multirow{2}{*}{$\rho_{\text{ATLAS}}(\FZ,\FL)$} \\
Measured value        &0.709  &0.299  &  \\ [\cmsTabSkip]
Uncertainty category  &  &  & \\ 
\hline
\multicolumn{4}{@{}l}{\textit{Samples size and background determination}} \\
\, \, \ Stat+bkg        & 0.012 & 0.008 & $-1.00$\\
\, \, \ Size of simulated samples    & 0.009 & 0.006 & $-1.00$\\ [\cmsTabSkip]
\multicolumn{4}{@{}l}{\textit{Detector modelling}} \\
\, \, \ JES                 & 0.005 & 0.003 & $-0.94$\\
\, \, \ JER                 & 0.006 & 0.003 & $-0.92$\\
\, \, \ JVF                 & 0.003 & 0.002 & $-0.99$\\
\, \, \ Jet reconstruction efficiency  & ${<}0.001$ & ${<}0.001$ & $-1.00$\\
\, \, \ Lepton efficiency   & 0.004 & 0.002 & $-0.99$\\
\, \, \ $\PQb$ tagging         & 0.002 & 0.001 & $-0.84$\\
\, \, \ Pileup              & \NA & \NA & \NA\\ [\cmsTabSkip]
\multicolumn{4}{@{}l}{\textit{Signal modelling}} \\
\, \, \ Top quark mass          & 0.002 & 0.007 & $-1.00$ \\
\, \, \ Simulation model choice      & 0.003 & 0.004 & 0.99 \\
\, \, \ Radiation and scales    & 0.003 & 0.006 & $-0.91$ \\
\, \, \ Top quark \pt           & \NA & \NA & \NA \\
\, \, \ PDF                     & 0.003 & 0.004 & $-1.00$ \\
\, \, \ Single top  method      & \NA & \NA & \NA \\ [\cmsTabSkip]
\multicolumn{4}{@{}l}{\textit{Total uncertainties}} \\
\, \, \ Systematic uncertainty & 0.014 & 0.013 & $-0.82$\\
\, \, \ Total uncertainty      & 0.019 & 0.015 & $-0.80$\\
\hline
\end{tabular}
\end{table}

\renewcommand{\arraystretch}{1.1}
\begin{table}[ht!]
\centering
\topcaption{Uncertainties in \FZ, \FL\, and their corresponding correlations  from the CMS $\Pe$+jets and $\Pgm$+jets measurements.
The uncertainty that is not applicable to this measurement, or which is included in other categories, is indicated by ``\NA''. The line ``Systematic uncertainty'' represents the quadratic sum of all the systematic uncertainty sources except for the uncertainties in the background determination and the integrated luminosity, which are included in the ``Stat+bkg'' category. The quoted correlation values are obtained via the procedures  described in Section~\ref{sec:correlation}.
\label{tab:uncertainties_CMSljets}
}
\cmsTable{
\begin{tabular}{@{}lcccccc@{}}
\hline
 & \multicolumn{3}{@{}c@{}}{CMS $\Pe$+jets} & \multicolumn{3}{@{}c@{}}{CMS $\Pgm$+jets} \\ 
        & \FZ & \FL &
        \multirow{2}{*}{$\rho_{\text{CMS}}^{\Pe\text{+jets}}(\FZ,\FL)$} &
        \FZ & \FL &
        \multirow{2}{*}{$\rho_{\text{CMS}}^{\Pgm\text{+jets}}(\FZ,\FL)$} \\
Measured value        &0.705  &0.304  & &0.685  &0.328  & \\ [\cmsTabSkip]
Uncertainty category  &       &       &      &       &       &      \\
\hline
\multicolumn{7}{@{}l}{\textit{Samples size and background determination}} \\
\, \, \ Stat+bkg        & 0.028   & 0.011   & $-0.87$   & 0.016  & 0.010   & $-0.88$  \\
\, \, \ Size of simulated samples    & 0.002   & 0.001   &  $-0.95$   & 0.002  & 0.001   &  $-0.96$  \\ [\cmsTabSkip]
\multicolumn{7}{@{}l}{\textit{Detector modelling}} \\
\, \, \ JES                 & 0.004   & 0.003   & $-1.00$   &  0.005  & 0.003   & $-1.00$  \\
\, \, \ JER                 & 0.001   & 0.002   & $-1.00$   & 0.004  & 0.003   & $-1.00$  \\
\, \, \ JVF                 & \NA   & \NA   & \NA   & \NA  & \NA   & \NA  \\
\, \, \ Jet reconstruction efficiency  & \NA   & \NA   & \NA   & \NA  & \NA   & \NA  \\
\, \, \ Lepton efficiency   & 0.001   & 0.002   & $-1.00$   & 0.001  & 0.001   & $-1.00$  \\
\, \, \ $\PQb$ tagging         & 0.001   & ${<}0.001$   & $-1.00$   & 0.001  & ${<}0.001$   & $-1.00$  \\
\, \, \ Pileup             & 0.001  & 0.001   & $-1.00$   & ${<}0.001$  & ${<}0.001$   & $-1.00$  \\ [\cmsTabSkip]
\multicolumn{7}{@{}l}{\textit{Signal modelling}} \\
\, \, \ Top quark mass          & 0.012   & 0.008   & $-0.99$   & 0.009  & 0.006   & $-1.00$  \\
\, \, \ Simulation model choice      & 0.015   & 0.010   & $-0.87$   & 0.008  & 0.004   & $0.20$  \\
\, \, \ Radiation and scales    & 0.007   & 0.005   & $-1.00$   & 0.014  & 0.006   & $-0.83$  \\
\, \, \ Top quark \pt          & 0.011   & 0.010   & $-1.00$   & ${<}0.001$  & 0.001   & $-1.00$  \\
\, \, \ PDF                     & 0.004   & 0.001   & $-0.92$   & 0.002  & 0.001   & $-0.15$  \\
\, \, \ Single top  method      & \NA   & \NA   & \NA   & \NA  & \NA   & \NA  \\ [\cmsTabSkip]
\multicolumn{7}{@{}l}{\textit{Total uncertainties}} \\
\, \, \ Systematic uncertainty  & 0.024   & 0.018   & $-0.93$   & 0.020  & 0.010   & $-0.71$ \\
\, \, \ Total uncertainty       & 0.037   & 0.021   & $-0.87$   & 0.025  & 0.014   & $-0.78$ \\
\hline
\end{tabular}
}
\end{table}

\renewcommand{\arraystretch}{1.1}
\begin{table}[ht!]
\centering
\topcaption{Uncertainties in \FZ, \FL\, and their corresponding correlations from the CMS (single top) measurement. The uncertainty that is not applicable to this measurement, or which is included in other categories, is indicated by ``\NA''. The line ``Systematic uncertainty'' represents the quadratic sum of all the systematic uncertainty sources except for the uncertainties in the background determination and the integrated luminosity, which are included in the ``Stat+bkg'' category. The quoted correlation values are obtained via the procedures described in Section~\ref{sec:correlation}.}
\label{tab:uncertainties_CMSst}
\begin{tabular}{@{}lccc@{}}
\hline
& \multicolumn{3}{@{}c@{}}{CMS (single top)} \\ 
                              & \FZ & \FL & \multirow{2}{*}{$\rho_{\text{CMS}}^{\text{st}}(\FZ,\FL)$} \\
Measured value        &0.720  &0.298  & \\ [\cmsTabSkip]
Uncertainty category  &  &  &  \\
\hline
\multicolumn{4}{@{}l}{\textit{Samples size  and background determination}} \\
\, \, \ Stat+bkg        & 0.041 & 0.031 & $-0.90$\\
\, \, \ Size of simulated samples    & 0.020 & 0.012 & $-0.96$\\
\multicolumn{4}{@{}l}{\textit{Detector modelling}} \\
\, \, \ JES                 & 0.004 & 0.004 & $-1.00$\\
\, \, \ JER                 & 0.001 & 0.001 & $-1.00$\\
\, \, \ JVF                 & \NA & \NA & \NA \\
\, \, \ Jet reconstruction efficiency  & \NA & \NA & \NA \\
\, \, \ Lepton efficiency   & ${<}0.001$ & ${<}0.001$ & $-1.00$ \\
\, \, \ $\PQb$ tagging         & 0.006 & 0.006 & $-1.00$\\
\, \, \ Pileup             & 0.003 & 0.003 & $-1.00$\\
\multicolumn{4}{@{}l}{\textit{Signal modelling}} \\
\, \, \ Top quark mass          & 0.005 & 0.007 & $-1.00$\\
\, \, \ Simulation model choice      & 0.002 & 0.003 & $-1.00$\\
\, \, \ Radiation and scales    & 0.023 & 0.019 & $-1.00$\\
\, \, \ Top quark \pt           & \NA & \NA & \NA \\
\, \, \ PDF                     & 0.004 & 0.004 & $-0.97$\\
\, \, \ Single top  method      & 0.012 & 0.015 & $-1.00$\\
\multicolumn{4}{@{}l}{\textit{Total uncertainties}} \\
\, \, \ Systematic uncertainty & 0.035 & 0.029 & $-0.96$ \\
\, \, \ Total uncertainty      & 0.054 & 0.043 & $-0.92$\\
\hline
\end{tabular}
\end{table}

\begin{figure}[h]
\centering
\includegraphics[width = 0.95\textwidth]{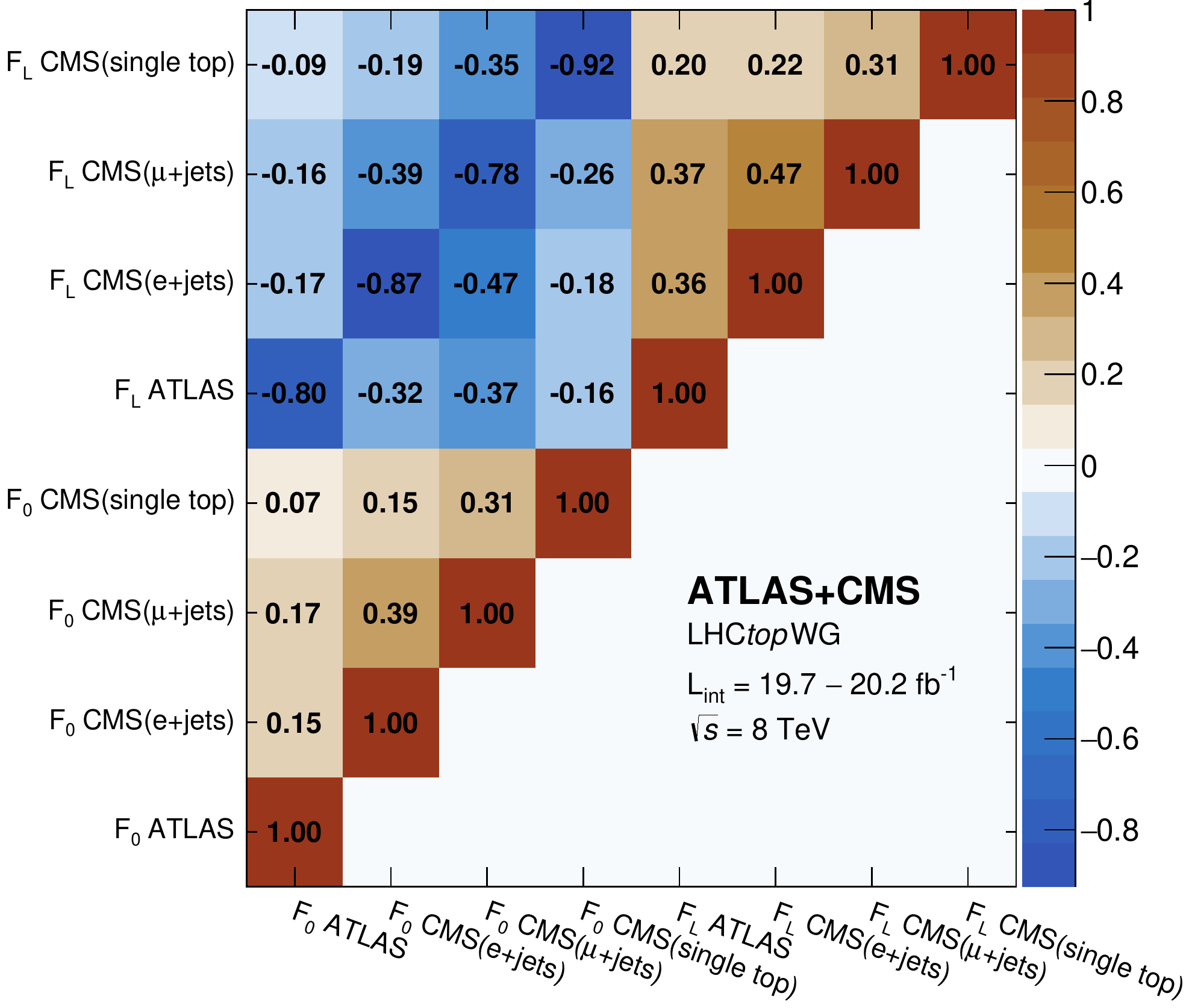}
\caption{The total correlation between the input measurements of the combination.
\label{fig:BlueCorrMatrix}
}
\end{figure}

Figure~\ref{fig:BlueCorrMatrix} presents the total correlation values between the input measurements. Typically, \FZ\ and \FL\ are highly anticorrelated within the same measurement. The three \ttbar measurements (ATLAS, CMS ($\Pe$+jets), and CMS ($\Pgm$+jets)) are also correlated or anti-correlated, with the absolute values of the correlations ranging around 30 to 40\%. The correlations of the CMS (single top) measurement with the  CMS ($\Pe$+jets) and CMS ($\Pgm$+jets) measurements are around 20\% in the absolute value, and are generally smaller with the ATLAS measurement.

\section{Results}
\label{sec:results}

The combination is performed by finding the best linear unbiased estimator (BLUE)~\cite{Lyons:1988rp,Valassi:2003mu} with the method implemented in Ref.~\cite{Nisius:2014wua}. The BLUE method finds the coefficients of the linear combination of the input measurements by minimizing the total uncertainty of the combined result, taking into account both the statistical and systematic uncertainties, as well as the correlations between the inputs.
 In this analysis, the measurements of \FZ\ and \FL\ are combined while \FR\ is obtained as \mbox{$\FR = 1-\FZ-\FL$}. As no further constraints on the observables were placed, values outside the range [0, 1] are allowed for the three polarization fractions. 
The total correlation between \FZ\ and \FL\ obtained from the combination is taken into account in the estimation of the uncertainty in the \FR\ value.

The results of the combination of the polarization fractions measurements are
\begin{equation*}
\begin{aligned}
  \FZ &=& \combFZ \pm \combFZStat\,\text{(stat+bkg)} \pm \combFZSyst\,\text{(syst)}, \\
  \FL &=& \combFL \pm \combFLStat\,\text{(stat+bkg)} \pm \combFLSyst\,\text{(syst)},
\end{aligned}
\end{equation*}
with a total correlation of \combRho. Using the unitarity
constraint on the polarization fractions, the fraction of events with a \wbos with right-handed polarization is calculated to be
\begin{equation*}
\begin{aligned}
\FR = \text{\combFR~}\pm {\combFRstat\,\text{(stat+bkg)} \pm \combFRsys\,\text{(syst)}},
\end{aligned}
\end{equation*}
where the first quoted uncertainty includes the statistical part and uncertainties in the background determination, and the second uncertainty refers to the remaining systematic contribution. From these results, an upper limit of $\FR < \text{\combFRLim}$ at 95\% confidence level (\CL) is set. The limit is set using the Feldman--Cousins method~\cite{refFC}, considering that \FR\ follows a normal distribution, and that it is physically bound to $\FR \geq 0$. The relative uncertainty on \FZ\ and \FL\ is 2.0 and 3.5\%, respectively, including systematic and statistical components.

\begin{figure}
\centering
\includegraphics[width = 1.0\textwidth]{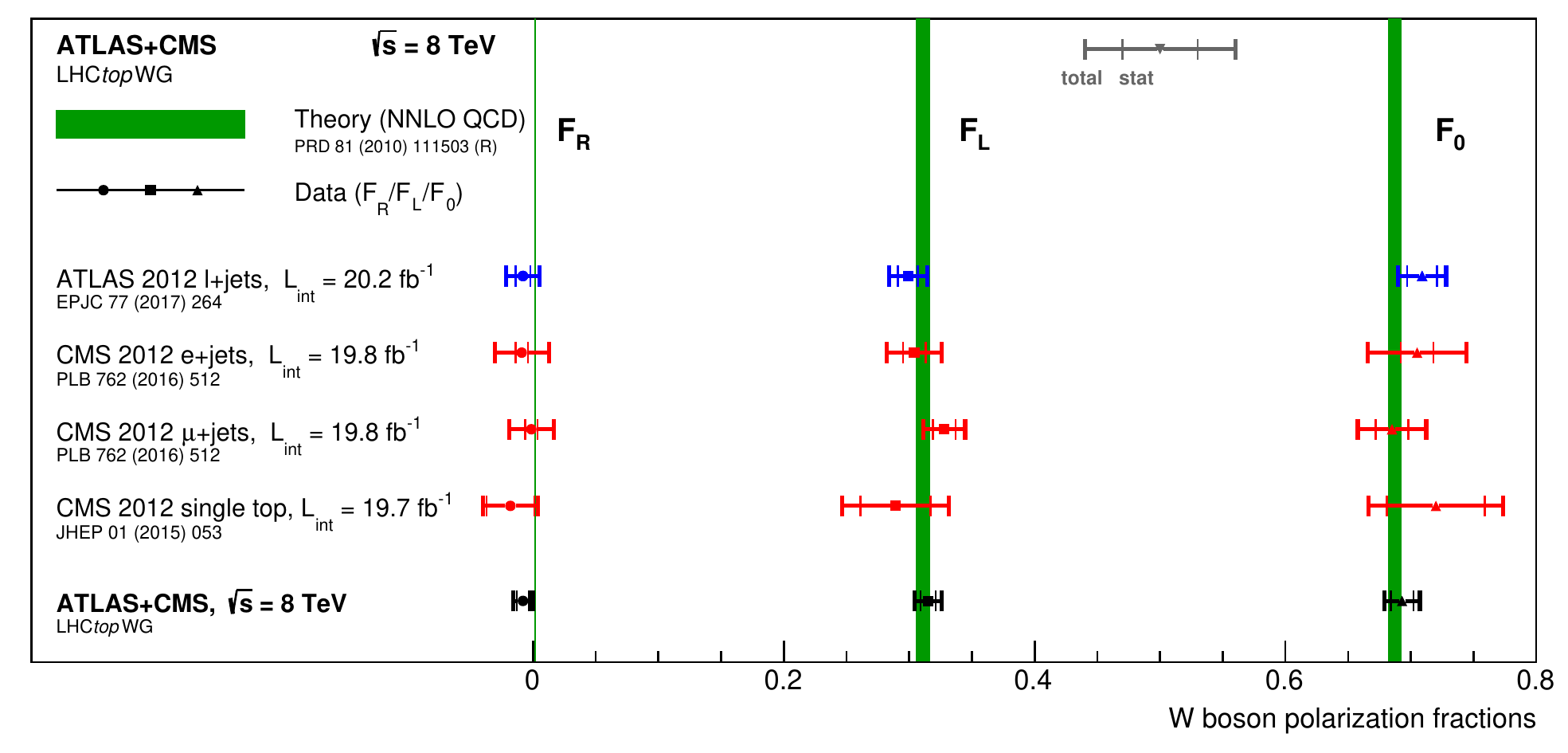}
\caption{Overview of the four measurements, as well as the results of the combination. The inner and outer error bars correspond to the statistical and the total uncertainties, respectively. The inner bars for the combination include also the background determination uncertainties. The vertical solid line indicates the predictions of NNLO QCD calculations~\cite{Czarnecki:2010gb}.
\label{fig:summary}}
\end{figure}

Figure~\ref{fig:summary} shows an overview of the four measurements included in the combination and the result of the combination together with the polarization fractions predicted by NNLO QCD calculations. The uncertainties in the NNLO predictions, presented with vertical bands, include an uncertainty of 1.3\GeV in the top quark mass, uncertainties in the $\PQb$ quark and \PW\ boson masses, and in \alpS. The combined \FR\ value is negative, as this is not explicitly forbidden in the combination, but compatible with the predictions within the uncertainties. The measurements are consistent with each other and with the NNLO QCD prediction.

The $\chi^2$ and upper tail probability of the combination are \combChi\ and \combProb\ respectively.  The combination includes four sets of measurements, each composed of two highly anticorrelated observables, and two fit parameters of the combination, \ie the combined \FZ\ and \FL. A detailed breakdown of the uncertainties is presented in Table~\ref{tab:uncertainties_LHC}.
The dominant uncertainties are those arising from the statistical uncertainty on data and background estimation (stat+bkg), followed by the uncertainties in the radiation and scales modelling, the limited size of the simulated samples, and simulation model choice. The total detector modelling uncertainty is minor, smaller than the uncertainties in the stat+bkg category. 
The measurement with the highest impact in the determination of \FZ~is ATLAS, while CMS ($\Pgm$+jets) dominates the combined \FL~determination. The impact of the CMS ($\Pe$+jets) and CMS ($\Pgm$+jets) measurements is not directly comparable to the other input measurements that already include the electron and muon channels together. As a test, the combination is repeated, using a pre-combined CMS ($\Pe$+jets) + CMS ($\Pgm$+jets) input, and the results are unchanged.
The ATLAS+CMS combined fractions and uncertainties are identical in both cases, with a small variation on the resulting (\FZ, \FL) correlation, being 1.5\% smaller for the cross-check combination.

In another test, the CMS (single top) measurement was removed from the combination. The impact on the combined fractions and uncertainties  is less than 1.5\%.

The combination yields an important improvement in precision, as compared to the most precise individual published measurements~\cite{TOPQ-2016-02,CMS-TOP-13-008}. Improvements of 25 and 29\% relative to the most precise single measurement are found for the precision of the combined measurements of \FZ\ and \FL, respectively.
The improvement is estimated with respect to the published values of the \wbos polarization fraction determination that is given in Table~\ref{tab:measurements}. The total correlation between the combined fractions is similar to those in the input measurements, and their uncertainties are smaller. These two factors lead to a combined right-handed polarization fraction \FR\ that is almost a factor two more precise than in previous publications.

\renewcommand{\arraystretch}{1.1}
\begin{table}[ht!]
\centering
\topcaption{
Results of the ATLAS and CMS combination: \wbos polarization fraction values and uncertainties. The combined \FZ\ and \FL\ values are anticorrelated, with $\rho$=\combRho.
\label{tab:uncertainties_LHC}
}
\begin{tabular}{@{}lcc@{}}
\hline
& \multicolumn{2}{@{}c@{}}{ATLAS+CMS combination} \\ 
           & \FZ  & \FL  \\
Fractions &   0.693 &  0.315 \\ 
Uncertainty category       &   &   \\
\hline
\multicolumn{3}{@{}l}{\textit{Samples size  and background determination}} \\
\, \, \ Stat+bkg  &    0.009 &  0.006 \\
\, \, \ Size of simulated samples &    0.005 &  0.003 \\ [\cmsTabSkip]
\multicolumn{3}{@{}l}{\textit{Detector modelling}} \\
\, \, \    JES &    0.004 &  0.002 \\
\, \, \    JER &    0.004 &  0.002 \\
\, \, \    JVF &    0.001 &  0.001 \\
\, \, \ Jet reconstruction &    ${<}0.001$ &  ${<}0.001$ \\
\, \, \ Lepton efficiency &  0.002 &  0.001 \\
\, \, \ $\PQb$ tagging &    0.001 &  0.001 \\
\, \, \ Pileup &    ${<}0.001$ &  ${<}0.001$ \\ [\cmsTabSkip]
\multicolumn{3}{@{}l}{\textit{Signal modelling}} \\
\, \, \ Top quark mass &    0.003 &  0.004 \\
\, \, \ Simulation model  choice &    0.006 &  0.005 \\
\, \, \ Radiation and scales  &    0.005 &  0.004 \\
\, \, \ Top quark \pt &    0.001 &  0.002 \\
\, \, \ PDF &    0.001 &  0.001 \\
\, \, \ Single top  method &    0.001 &  ${<}0.001$ \\ [\cmsTabSkip]
Total uncertainty            &    0.014 &  0.011 \\
\hline
\end{tabular}
\end{table}

\subsection{Stability tests}
\label{subsec:stability}
The hypotheses assumed for the correlations between the measurements, as defined in Sections~\ref{sec:correlation} and~\ref{sec:partialcorrel}, are based on the best knowledge of the similarities and differences in the detectors, analysis methods, and simulations used in each measurement. Nevertheless, some of these correlations cannot be precisely determined. The checks described in this section are performed to test the stability of the results against this potential lack of knowledge.

\noindent{\textit{$\rho_{\text{LHC}}(\Fi,\Fi)$~hypothesis (with $i$=0,L) for the JES uncertainty:}}
The correlation value $\rho_{\text{LHC}}(\Fi,\Fi)=0.2$ was estimated according to the prescription given in Ref.~\cite{ATL-PHYS-PUB-2015-049} and the description in Section~\ref{sec:partialcorrel}. The impact of this assumption is evaluated by repeating the combination by varying $\rho_{\text{LHC}}(\Fi,\Fi)$ in the interval between 0.0 and 0.4, in steps of 0.1. The fraction values and uncertainties remained  unchanged in the entire probed range. The $\chi^2$ of the fit, the probability, and the total (\FZ,\FL) correlation are found to be stable with a relative shift of less than 0.5\%. 

\noindent{\textit{$\rho_{\text{LHC}}(\Fi,\Fi)$ and $\rho_{\text{CMS}}^{\text{st,}\ell\text{+jets}}(\Fi,\Fi)$ hypotheses for the radiation and scales uncertainties:}}
Although addressing similar effects, the radiation and scales uncertainties are estimated in three different ways for ATLAS, CMS (single top), and the other CMS measurements, with different levels of correlations among them. Therefore, the two hypotheses, $\rho_{\text{LHC}}(\Fi,\Fi)=0.5$ and $\rho_{\text{CMS}}^{\text{st,}\ell\text{+jets}}(\Fi,\Fi)=1$, are tested simultaneously,  by variation in steps of 0.1 in the interval between 0 and 0.5 for $\rho_{\text{LHC}}(\Fi,\Fi)$ and between 0.6 and 1.0 for $\rho_{\text{CMS}}^{\text{st,}\ell\text{+jets}}(\Fi,\Fi)$. The resulting polarization fraction mean values and uncertainties remained unchanged in the whole ranges. Small variations, below the percent level, are observed for the total correlation and fit probability.

\noindent{\textit{JES versus radiation and scales correlations:}}
Since the JES and radiation and scales uncertainties are among the dominant sources of uncertainty with significant correlation between measurements, an additional test was performed varying the two correlation hypotheses simultaneously, rather than separately. The results of this test also show stable combination with maximum relative shifts of about 2\% for the $\chi^2$ and probability and about 0.6\% for the total correlation. The combined fractions and uncertainties are found to be stable, with negligible variations for all probed hypotheses.

\noindent{\textit{$\rho_{\text{CMS}}^{\Pe,\Pgm\text{+jets}}(\Fi,\Fi)$ and $\rho_{\text{CMS}}^{\text{st,}\ell\text{+jets}}(\Fi,\Fi)$ hypothesis for statistical+background uncertainty:}}   Small correlations that could arise from the background modelling from simulated samples are
neglected in the combination by assuming $\rho_{\text{CMS}}^{\Pe,\Pgm\text{+jets}}(\Fi,\Fi)=0$ and $\rho_{\text{CMS}}^{\text{st,}\ell\text{+jets}}(\Fi,\Fi)=0$. In order to investigate the effect of these hypotheses, the combination was repeated by varying $\rho_{\text{CMS}}^{\Pe,\Pgm\text{+jets}}(\Fi,\Fi)$ and $\rho_{\text{CMS}}^{\text{st,}\ell\text{+jets}}(\Fi,\Fi)$, using for both the same correlation values in the range [0.0, 0.7] in steps of 0.1. In the interval between 0.0 and 0.6, the fraction values are varied by a maximum of 1.3\%, with \FZ\ going from 0.693 to 0.687, and \FL\ from 0.314 to 0.319. At 0.7, the combination yields $\FZ=0.684 \pm 0.014$ and $\FL=0.321 \pm 0.010$, which is the maximum variation observed in all tests performed in this study. However, in this case the fit probability  decreases to 28\%, suggesting that the correlation assumption of 0.7 is less favoured.  The fit combination does not converge for  unreasonable values, \ie correlation values above 0.7. 

In conclusion, the tests reported in this section indicate that the combined results are robust against variations of some poorly known or unknown input correlations. The correlations are varied over a large range, and in all cases the observed deviation from the nominal results are well covered by the  uncertainties in the combined result.

\subsection{Limits on anomalous couplings}
\label{sec:anomcoupl}
The result of the combination of the polarization fractions measurements can be used to set limits on  beyond-the-SM physics contributing to the \wtb vertex. In the two approaches presented in this section, only new physics contributions to the top quark decay vertex are considered---effects at the production vertex in single top quark processes are disregarded.

\begin{figure}[htb!]
\centering
\includegraphics[width = 0.49\textwidth]{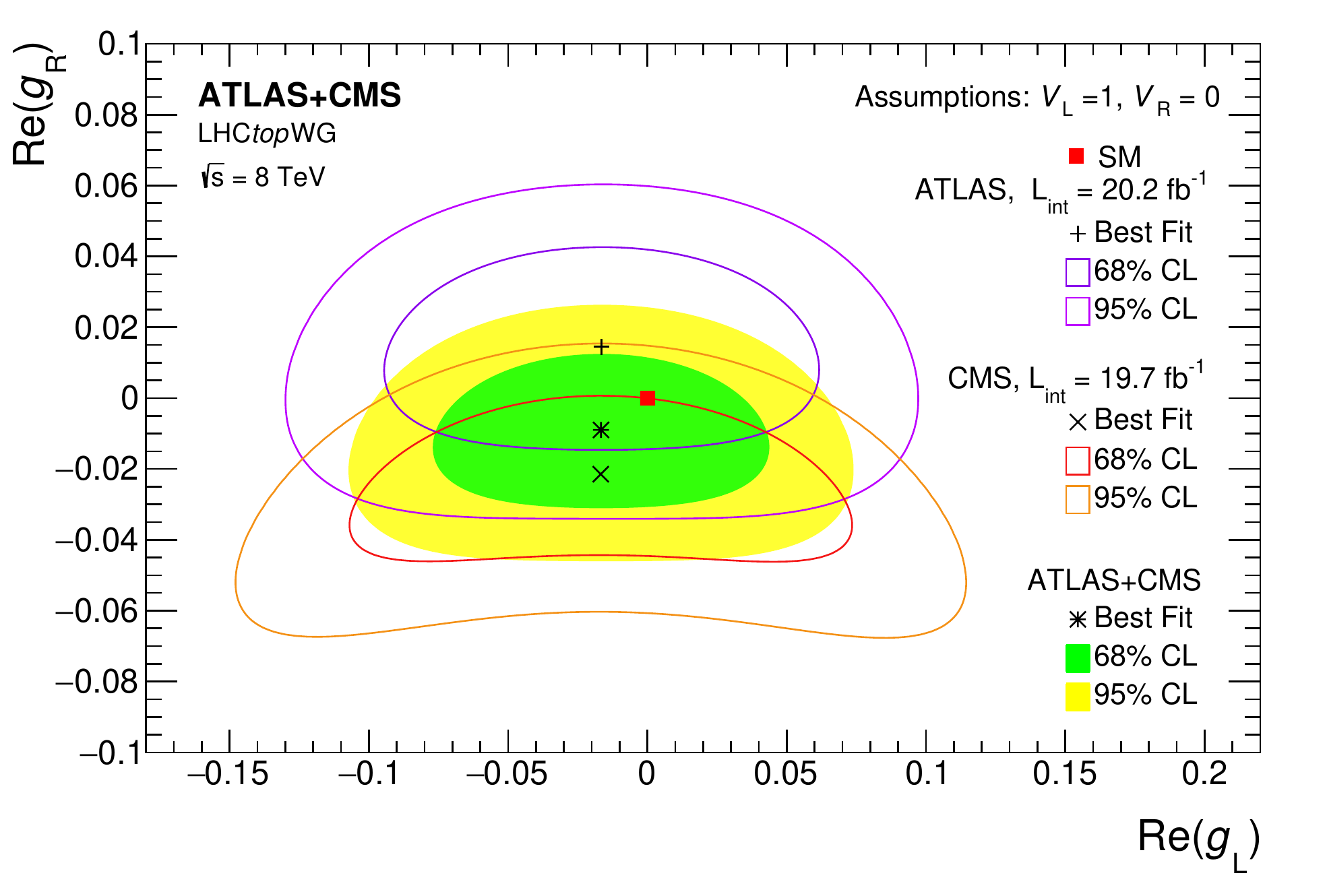}
\includegraphics[width = 0.49\textwidth]{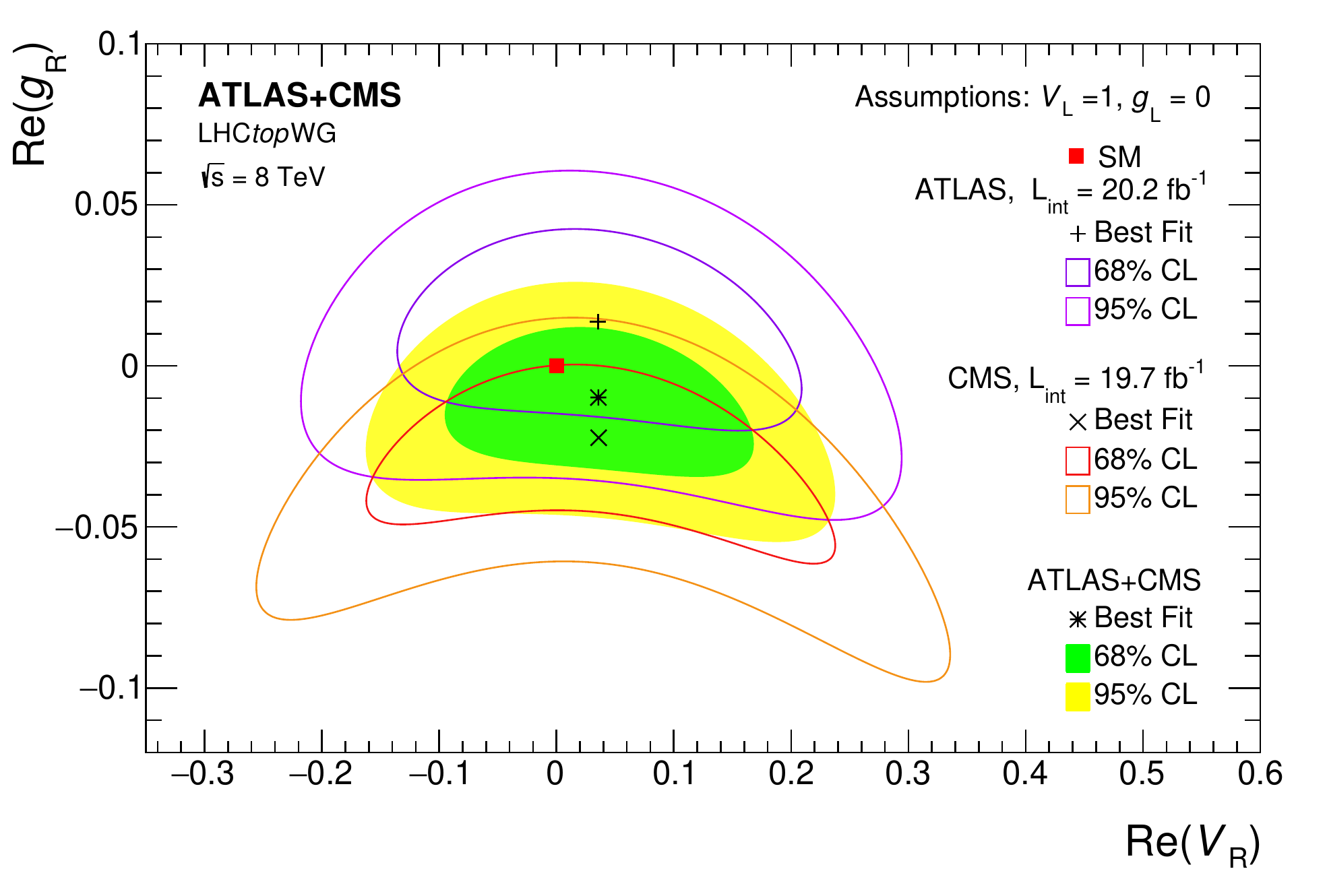}
\caption{Allowed regions for the \wtb anomalous (left) left- and right-handed tensor couplings, and (right) right-handed vector and tensor coupling. The limits are obtained from the ATLAS, CMS, and the combined measurements of the \wbos polarization fractions at 68 and 95\% \CL. The limits from CMS are obtained using the pre-combined result of all CMS input measurements. The anomalous couplings are assumed to be real.}
\label{fig:2D_limits}
\end{figure}

In a first approach, the structure of the \wtb vertex is parameterized in a general form in effective field theory, expanding the SM Lagrangian to include dimension-six terms
\begin{equation}
\begin{aligned}
\mathcal{L}_{\text{\wtb}} = - \frac{g}{\sqrt 2} \PAQb \, \gamma^{\mu} \left( V_{\mathrm{L}} P_{\mathrm{L}}
  + V_{\mathrm{R}} P_{\mathrm{R}} \right) \PQt\; \PW_\mu^-  - \frac{g}{\sqrt 2} \PAQb \, \frac{i \sigma^{\mu \nu} \PQq_\nu}{m_\PW}
  \left( g_{\mathrm{L}} P_{\mathrm{L}} + g_{\mathrm{R}} P_{\mathrm{R}} \right) \PQt\; \PW_\mu^- + \mathrm{h.c.},
\label{eq:Wtb}
\end{aligned}
\end{equation}
where $V_{\text{L,R}}$ and $g_{\text{L,R}}$ are left- and right-handed vector and tensor couplings, respectively. Here, $P_{\text{L,R}}$ refers to the left- and right-handed chirality projection operators, $m_\PW$ to the \wbos mass, and $g$ to the weak coupling constant, as detailed in Refs.~\cite{Buchmuller:1985jz,AguilarSaavedra:2008zc}.
In the SM, $V_{\text{L}}$ is given by the Cabibbo--Kobayashi--Maskawa (CKM) matrix element $V_{\text{tb}}$, with a measured value of $\approx 1$, while $V_{\text{R}}=g_{\text{L}}=g_{\text{R}}=0$ at the tree level. Using this formalism, the polarization fractions can be translated into the couplings
$V_{\mathrm{L}},\ V_{\mathrm{R}},\ g_{\mathrm{L}}$, and $\ g_{\mathrm{R}}$ (as discussed \eg in Ref.~\cite{AguilarSaavedra:2006fy}).
The two independent \wbos polarization measurements, \FZ\ and \FL, cannot fully constrain the four \wtb couplings. Therefore additional assumptions have to be made.
Figure~\ref{fig:2D_limits} shows the limits on the left- and right-handed tensor couplings, while the  other couplings are fixed to their SM values, as well as limits on the right-handed vector and tensor couplings,  with the other couplings fixed to their SM values. Limits on these anomalous couplings are set using the \textsc{EFT\textit{fitter}} tool~\cite{Castro:2016jjv}. The anomalous couplings are assumed to introduce no additional CP violation, and are taken to be real.
 The allowed regions at 68 and 95\% \CL and the most probable couplings values are shown, as derived from the measured polarization fractions reported in Refs.~\cite{TOPQ-2016-02, CMS-TOP-13-008}, and from the combined results presented in this paper. A second region allowed by the \wbos polarization measurements around ${\text{Re}}(g_{\mathrm{R}})=0.8$ is excluded by the single top quark cross section measurements~\cite{CMS-st,TOPQ-2015-05}, and therefore is not shown in this figure. Table~\ref{tab:1d_limits} shows the 95\% \CL intervals for each anomalous coupling, while fixing all others to their SM values. These limits correspond to the set of smallest intervals containing 95\% of the marginalized posterior distribution for the corresponding parameter.

\begin{table}[!ht]
  \centering
  \topcaption{
Allowed ranges for the anomalous couplings $V_{\text{R}}$, $g_{\text{L}}$, and $g_{\text{R}}$ at 95\% \CL. The limit on each coupling is obtained while fixing all other couplings to their SM value. 
The limits from CMS are obtained using the pre-combined result of all CMS input measurements. The anomalous couplings are assumed to be real.
}
  \begin{tabular}{c cc l}
\hline
 &\multicolumn{3}{@{}c}{95\% \CL interval}\\
Coupling & ATLAS & CMS  & ATLAS+CMS combination\\
\hline
  Re($V_{\text{R}}$) & $[-0.17, 0.25]$ & $[-0.12, 0.16]$ & $[-0.11, 0.16]$\\
Re($g_{\text{L}}$) & $[-0.11, 0.08]$ & $[-0.09, 0.06]$ & $[-0.08, 0.05]$\\
Re($g_{\text{R}}$) & $[-0.03, 0.06]$ & $[-0.06, 0.01]$ & $[-0.04, 0.02]$\\
\hline
\end{tabular}
\label{tab:1d_limits}
\end{table}

In a similar way, limits are set in terms of Wilson coefficients.
In this second approach, effects of beyond-the-SM physics at a high scale $\Lambda$ are described by an effective Lagrangian~\cite{EFTlhc,Burges:1983zg, Leung:1984ni, Buchmuller:1985jz,Zhang:2017} as
\begin{equation}
\begin{aligned}
{\mathcal-{L}}^{\text{eff}} = {\mathcal{L}}^{\text{SM}} + \Sigma_x \frac{C_x}{\Lambda^2}O_x + {\mathcal{O}}\left(\frac{1}{\Lambda^3}\right) + \cdots
\label{eq:LagEff}
\end{aligned}
\end{equation}
where $O_x$ are dimension-six gauge-invariant operators and $C_x$ are the complex constants known as Wilson coefficients that give the strength of the corresponding operator. Only dimension-six operators are considered in this analysis. The relevant operators affecting the general effective \wtb vertex can be found, \eg in Ref.~\cite{Zhang:2017}.
Three of these operators are of particular interest, since the measurement of the \wbos polarization is able to constrain their corresponding Wilson coefficients. These operators are:
\begin{linenomath}
\begin{equation}
\begin{aligned}
O_{\phi \phi} &= {\text{i}}(\Tilde{\phi}^{\dagger} D_\mu \phi) (\PAQt_{\text{R}} \gamma^\mu \PQb_{\text{R}}), \\
O_{\PQt \PW}  &= (\PAQq_{\text{L}} \sigma^{\mu\nu} \tau^I \PQt_{\text{R}}) \Tilde{\phi} \PW^I_{\mu\nu}, \quad  \text{and}\\
O_{\PQb \PW}  &= (\PAQq_{\text{L}} \sigma^{\mu\nu} \tau^I \PQb_{\text{R}})\phi \PW^I_{\mu\nu},
\end{aligned}
\end{equation}
\end{linenomath}
where $\phi$ represents a weak doublet of the Higgs field, $\PQt_{\text{R}}$ and $\PQb_{\text{R}}$ are the weak singlets of the right-handed top and bottom quark fields, $\PQq^{\text{T}}_{\text{L}}=(\PQt,\PQb)_{\text{L}}$ denotes the $SU(2)_{\text{L}}$ weak doublet of the third generation left-handed quark fields, and $\tau^I$ is the usual Pauli matrix. Assuming the Wilson coefficients to be real, they can be trivially parameterized as functions of the anomalous couplings of Eq.~(\ref{eq:Wtb}) (as shown \eg in Refs.~\cite{Zhang:2017,AguilarSaavedra:2008zc}), thus, as functions of the \PW~polarization fractions.
The limits on each Wilson coefficient are derived from the measured fractions, as done for the anomalous couplings, fixing all others to their SM value, \ie to zero. They are shown at 95\% \CL in Table~\ref{tab:1d_limits_WC}.

\begin{table}[!ht]
  \centering
  \topcaption{
Allowed ranges for the Wilson coefficients $C_{\phi \phi}^{\ast}$, $C_{\PQb\PW}^{\ast}$, and $C_{\PQt\PW}$ at 95\% \CL. The limit on each coefficient is obtained while fixing all other coefficients to their SM values. The limits from CMS are obtained using the pre-combined result of all CMS input measurements. The numerical values are obtained by setting the $\Lambda$ scale to 1\TeV, and the coefficients are assumed to be real.}
  \begin{tabular}{c cc l}
\hline
 &\multicolumn{3}{@{}c}{95\% \CL interval}\\
Coefficient & ATLAS & CMS & ATLAS+CMS combination\\
\hline
  $C_{\phi \phi}^{\ast}$ & $[-5.64, 7.68]$ & $[-3.84, 4.92]$ & $[-3.48, 5.16]$\\
$C_{\PQb\PW}^{\ast}$ & $[-1.30, 0.96]$ & $[-1.06, 0.72]$ & $[-0.96, 0.67]$\\
$C_{\PQt\PW}$ & $[-0.34, 0.67]$ & $[-0.62, 0.19]$ & $[-0.48, 0.29]$\\
\hline
\end{tabular}

\label{tab:1d_limits_WC}
\end{table}

\section{Summary}
\label{sec:conclusion}

The combination of measurements of the \wbos polarization in top quark decays performed by the ATLAS and CMS Collaborations is presented.
The measurements are based on proton-proton collision data produced at the LHC at a centre-of-mass energy of 8\TeV, and corresponding to an integrated luminosity of about 20\fbinv for each experiment. The fractions of \wboss\ with longitudinal (\FZ) and left-handed (\FL) polarizations were measured in events containing a single lepton and multiple jets, enhanced in \ttbar or single top quark production processes. The results of the combination are
\begin{equation*}
\begin{aligned}
  \FZ = \combFZ \pm \combFZStat\,\text{(stat+bkg)} \pm \combFZSyst\,\text{(syst)}, \\
  \FL = \combFL \pm \combFLStat\,\text{(stat+bkg)} \pm \combFLSyst\,\text{(syst)},
\end{aligned}
\end{equation*}
where ``stat+bkg'' stands for the sum of the statistical and background determination uncertainties, and ``syst'' for the remaining systematic uncertainties. The fraction of \wboss\ with right-handed polarization, \FR , is estimated assuming that the sum of all polarization fractions equals unity, and by taking into account the correlation coefficient of the combination, \combRho. This leads to
\begin{equation*}
\begin{aligned}
    \FR = \text{\combFR~} \pm \combFRstat\,\text{(stat+bkg)} \pm \combFRsys\,\text{(syst)},
\end{aligned}
\end{equation*}
which corresponds to $\FR<\text{\combFRLim}$ at 95\% confidence level.

The results are consistent with the standard model predictions at next-to-next-to-leading-order precision in perturbative quantum chromodynamics. A limit on each anomalous \wtb coupling is set while fixing all others to their standard model values, with the allowed regions being $[-0.11, 0.16]$ for $V_{\text{R}}$, $[-0.08, 0.05]$ for $g_{\text{L}}$, and $[-0.04, 0.02]$ for $g_{\text{R}}$, at 95\% confidence level. All couplings are assumed to be real. Limits on Wilson coefficients are also derived in a similar manner.

\begin{acknowledgments}

We congratulate our colleagues in the CERN accelerator departments for the excellent performance of the LHC and thank the technical and administrative staffs at CERN and at other institutes for their contributions to the success of the ATLAS and CMS efforts.

We acknowledge the support of ANPCyT, Argentina; YerPhI, Armenia; ARC, Australia; BMWFW and FWF, Austria; ANAS, Azerbaijan; SSTC, Belarus; CNPq and FAPESP, Brazil; NSERC, NRC and CFI, Canada; CERN; CONICYT, Chile; CAS, MOST and NSFC, China; COLCIENCIAS, Colombia; MSMT CR, MPO CR and VSC CR, Czech Republic; DNRF and DNSRC, Denmark; IN2P3-CNRS, CEA-DRF/IRFU, France; SRNSFG, Georgia; BMBF, HGF, and MPG, Germany; GSRT, Greece; RGC, Hong Kong SAR, China; ISF and Benoziyo Center, Israel; INFN, Italy; MEXT and JSPS, Japan; CNRST, Morocco; NWO, Netherlands; RCN, Norway; MNiSW and NCN, Poland; FCT, Portugal; MNE/IFA, Romania; MES of Russia and NRC KI, Russian Federation; JINR; MESTD, Serbia; MSSR, Slovakia; ARRS and MIZ\v{S}, Slovenia; DST/NRF, South Africa; MINECO, Spain; SRC and Wallenberg Foundation, Sweden; SERI, SNSF and Cantons of Bern and Geneva, Switzerland; MOST, Taiwan; TAEK, Turkey; STFC, United Kingdom; DOE and NSF, United States of America. In addition, individual groups and members have received support from BCKDF, CANARIE, Compute Canada and CRC, Canada;  ERC, ERDF, Horizon 2020, Marie Sk{\l}odowska-Curie Actions and COST, European Union; Investissements d'Avenir Labex and Idex, ANR, France; DFG and AvH Foundation, Germany; Herakleitos, Thales and Aristeia programmes co-financed by EU-ESF and the Greek NSRF, Greece; BSF-NSF and GIF, Israel; CERCA Programme Generalitat de Catalunya and PROMETEO Programme Generalitat Valenciana, Spain; G\"{o}ran Gustafssons Stiftelse, Sweden; The Royal Society and Leverhulme Trust, United Kingdom.

We acknowledge the enduring support for the construction and operation of the LHC and the CMS detector provided by the following funding agencies: BMBWF and FWF (Austria); FNRS and FWO (Belgium); CNPq, CAPES, FAPERJ, FAPERGS, and FAPESP (Brazil); MES (Bulgaria); CERN; CAS, MoST, and NSFC (China); COLCIENCIAS (Colombia); MSES and CSF (Croatia); RPF (Cyprus); SENESCYT (Ecuador); MoER, ERC IUT, PUT and ERDF (Estonia); Academy of Finland, MEC, and HIP (Finland); CEA and CNRS/IN2P3 (France); BMBF, DFG, and HGF (Germany); GSRT (Greece); NKFIA (Hungary); DAE and DST (India); IPM (Iran); SFI (Ireland); INFN (Italy); MSIP and NRF (Republic of Korea); MES (Latvia); LAS (Lithuania); MOE and UM (Malaysia); BUAP, CINVESTAV, CONACYT, LNS, SEP, and UASLP-FAI (Mexico); MOS (Montenegro); MBIE (New Zealand); PAEC (Pakistan); MSHE and NSC (Poland); FCT (Portugal); JINR (Dubna); MON, RosAtom, RAS, RFBR, and NRC KI (Russia); MESTD (Serbia); SEIDI, CPAN, PCTI, and FEDER (Spain); MOSTR (Sri Lanka); Swiss Funding Agencies (Switzerland); MST (Taipei); ThEPCenter, IPST, STAR, and NSTDA (Thailand); TUBITAK and TAEK (Turkey); NASU (Ukraine); STFC (United Kingdom); DOE and NSF (USA).\\

\hyphenation{Rachada-pisek} Individuals have received support from the Marie-Curie programme and the European Research Council and Horizon 2020 Grant, contract Nos.\ 675440, 752730, and 765710 (European Union); the Leventis Foundation; the A.P.\ Sloan Foundation; the Alexander von Humboldt Foundation; the Belgian Federal Science Policy Office; the Fonds pour la Formation \`a la Recherche dans l'Industrie et dans l'Agriculture (FRIA-Belgium); the Agentschap voor Innovatie door Wetenschap en Technologie (IWT-Belgium); the F.R.S.-FNRS and FWO (Belgium) under the ``Excellence of Science -- EOS" -- be.h project n.\ 30820817; the Beijing Municipal Science \& Technology Commission, No. Z191100007219010; the Ministry of Education, Youth and Sports (MEYS) of the Czech Republic; the Deutsche Forschungsgemeinschaft (DFG) under Germany's Excellence Strategy -- EXC 2121 ``Quantum Universe" -- 390833306; the Lend\"ulet (``Momentum") Programme and the J\'anos Bolyai Research Scholarship of the Hungarian Academy of Sciences, the New National Excellence Program \'UNKP, the NKFIA research grants 123842, 123959, 124845, 124850, 125105, 128713, 128786, and 129058 (Hungary); the Council of Science and Industrial Research, India; the HOMING PLUS programme of the Foundation for Polish Science, cofinanced from European Union, Regional Development Fund, the Mobility Plus programme of the Ministry of Science and Higher Education, the National Science Center (Poland), contracts Harmonia 2014/14/M/ST2/00428, Opus 2014/13/B/ST2/02543, 2014/15/B/ST2/03998, and 2015/19/B/ST2/02861, Sonata-bis 2012/07/E/ST2/01406; the National Priorities Research Program by Qatar National Research Fund; the Ministry of Science and Education, grant no. 14.W03.31.0026 (Russia); the Tomsk Polytechnic University Competitiveness Enhancement Program and ``Nauka" Project FSWW-2020-0008 (Russia); the Programa Estatal de Fomento de la Investigaci{\'o}n Cient{\'i}fica y T{\'e}cnica de Excelencia Mar\'{\i}a de Maeztu, grant MDM-2015-0509 and the Programa Severo Ochoa del Principado de Asturias; the Thalis and Aristeia programmes cofinanced by EU-ESF and the Greek NSRF; the Rachadapisek Sompot Fund for Postdoctoral Fellowship, Chulalongkorn University and the Chulalongkorn Academic into Its 2nd Century Project Advancement Project (Thailand); the Kavli Foundation; the Nvidia Corporation; the SuperMicro Corporation; the Welch Foundation, contract C-1845; and the Weston Havens Foundation (USA). 

In addition, we gratefully acknowledge the computing centres and personnel of the Worldwide LHC Computing Grid for delivering so effectively the computing infrastructure essential to our analyses. In particular, the support from CERN, the ATLAS Tier-1 facilities at TRIUMF (Canada), NDGF (Denmark, Norway, Sweden), CC-IN2P3 (France), KIT/GridKA (Germany), INFN-CNAF (Italy), NL-T1 (Netherlands), PIC (Spain), ASGC (Taiwan), RAL (UK) and BNL (USA), the Tier-2 facilities worldwide and large non-WLCG resource providers is acknowledged gratefully. Major contributors of ATLAS computing resources are listed in Ref.~\cite{ATL-GEN-PUB-2016-002}.

\end{acknowledgments}

\bibliography{auto_generated}

\cleardoublepage\appendix \section{The CMS Collaboration \label{app:collab}}\begin{sloppypar}\hyphenpenalty=5000\widowpenalty=500\clubpenalty=5000\vskip\cmsinstskip
\textbf{Yerevan Physics Institute, Yerevan, Armenia}\\*[0pt]
A.M.~Sirunyan$^{\textrm{\dag}}$, A.~Tumasyan
\vskip\cmsinstskip
\textbf{Institut f\"{u}r Hochenergiephysik, Wien, Austria}\\*[0pt]
W.~Adam, F.~Ambrogi, T.~Bergauer, M.~Dragicevic, J.~Er\"{o}, A.~Escalante~Del~Valle, M.~Flechl, R.~Fr\"{u}hwirth\cmsAuthorMark{1}, M.~Jeitler\cmsAuthorMark{1}, N.~Krammer, I.~Kr\"{a}tschmer, D.~Liko, T.~Madlener, I.~Mikulec, N.~Rad, J.~Schieck\cmsAuthorMark{1}, R.~Sch\"{o}fbeck, M.~Spanring, W.~Waltenberger, C.-E.~Wulz\cmsAuthorMark{1}, M.~Zarucki
% ...
\vskip\cmsinstskip
\dag: Deceased\\
1:  Also at Vienna University of Technology, Vienna, Austria\\
% ...
\end{sloppypar}
\section{The ATLAS Collaboration \label{app:ATLAScollab}}\begin{sloppypar}\hyphenpenalty=5000\widowpenalty=500\clubpenalty=5000
 
\begin{flushleft}
\hypersetup{urlcolor=black}
{\Large The ATLAS Collaboration}

\bigskip

\AtlasOrcid[0000-0002-6665-4934]{G.~Aad}$^\textrm{\scriptsize 102}$,    
\AtlasOrcid[0000-0002-5888-2734]{B.~Abbott}$^\textrm{\scriptsize 128}$,    
\AtlasOrcid{D.C.~Abbott}$^\textrm{\scriptsize 103}$,    
\AtlasOrcid[0000-0002-2788-3822]{A.~Abed~Abud}$^\textrm{\scriptsize 36}$,    
\AtlasOrcid[0000-0002-1002-1652]{K.~Abeling}$^\textrm{\scriptsize 53}$,    
\AtlasOrcid[0000-0002-2987-4006]{D.K.~Abhayasinghe}$^\textrm{\scriptsize 94}$,    
\AtlasOrcid[0000-0002-8496-9294]{S.H.~Abidi}$^\textrm{\scriptsize 166}$,    
\AtlasOrcid[0000-0002-8279-9324]{O.S.~AbouZeid}$^\textrm{\scriptsize 40}$,    
\AtlasOrcid{N.L.~Abraham}$^\textrm{\scriptsize 155}$,    
\AtlasOrcid[0000-0001-5329-6640]{H.~Abramowicz}$^\textrm{\scriptsize 160}$
% ...
\bigskip
\\

% ...
$^{36}$CERN, Geneva; Switzerland.\\
% ...
$^{40}$Niels Bohr Institute, University of Copenhagen, Copenhagen; Denmark.\\
% ...
$^{53}$II. Physikalisches Institut, Georg-August-Universit\"{a}t G\"ottingen, G\"ottingen; Germany.\\
% ...
$^{94}$Department of Physics, Royal Holloway University of London, Egham; United Kingdom.\\
% ...
$^{102}$CPPM, Aix-Marseille Universit\'e, CNRS/IN2P3, Marseille; France.\\
$^{103}$Department of Physics, University of Massachusetts, Amherst MA; United States of America.\\
% ...
$^{128}$Homer L. Dodge Department of Physics and Astronomy, University of Oklahoma, Norman OK; United States of America.\\
% ...
$^{155}$Department of Physics and Astronomy, University of Sussex, Brighton; United Kingdom.\\
% ...
$^{160}$Raymond and Beverly Sackler School of Physics and Astronomy, Tel Aviv University, Tel Aviv; Israel.\\
% ...
$^{166}$Department of Physics, University of Toronto, Toronto ON; Canada.\\
% ...

\end{flushleft}

\end{sloppypar}
\end{document}